\documentclass[useAMS,usenatbib]{mn2e}

\usepackage[pdftex,pdfpagemode={UseOutlines},bookmarks,bookmarksopen,colorlinks,linkcolor={blue},citecolor={blue},urlcolor={red}]{hyperref}
\usepackage{float}
\usepackage{graphicx}
\usepackage{caption}
\usepackage{subcaption}
\usepackage{natbib}
\usepackage{rotating,longtable,lscape}

\title[The atmosphere of WASP-80b]{WASP-80b has a dayside within the T-dwarf range\thanks{using data acquired with the {\it Spitzer} Space Telescope (PID 90159), and ground-based data collected at ESO's La Silla Observatory, Chile: HARPS on the ESO 3.6m (Prog ID 089.C-0151 \& 091.C-0184), the Swiss {\it Euler} Telescope, and TRAPPIST. The data is publicly available at the \textit{CDS} Strasbourg and on demand to the main author.}}
\author[Amaury H.~M.~J. Triaud et al.]
{Amaury H.~M.~J. Triaud$^{1,2,3}$\thanks{E-mail: triaud@mit.edu}, 
Micha\"el Gillon$^{4}$,
David Ehrenreich$^{5}$,
Enrique Herrero$^{6}$,
\newauthor
Monika Lendl$^{4,5}$,
David R. Anderson$^{7}$,
Andrew Collier Cameron$^{8}$,
Laetitia Delrez$^{4}$,
\newauthor
Brice-Olivier Demory$^{9}$,
Coel Hellier$^{7}$,
Kevin Heng$^{10}$,
Emmanuel Jehin$^{4}$,
Pierre F.~L. Maxted$^{7}$,
Don Pollacco$^{11}$,
\newauthor
Didier Queloz$^{9}$,
Ignasi Ribas$^{6}$,
Barry Smalley$^{7}$,
Alexis M.~S. Smith$^{12}$,
St\'ephane Udry$^{5}$\\
$^{1}$Centre for Planetary Sciences, University of Toronto at Scarborough, 1265 Military Trail, Toronto, ON, M1C 1A4, Canada\\
$^{2}$Department of Astronomy \& Astrophysics, University of Toronto, Toronto, Ontario M5S 3H4, Canada\\
$^{3}$Kavli Institute for Astrophysics \& Space Research, Massachusetts Institute of Technology, Cambridge, MA 02139, USA\\
$^{4}$Institut d'Astrophysique et de G\'eophysique, Universit\'e de Li\`ege, All\'ee du 6 Ao\^ut 17, Sart Tilman, 4000 Li\`ege 1, Belgium\\
$^{5}$Observatoire Astronomique de l'UniversitŽ\'e de Gen\`eve, Chemin des Maillettes 51, CH-1290 Sauverny, Switzerland\\
$^{6}$Institut de Ci\`encies de l'Espai (CSIC-IEEC), Campus UAB, Facultat de Ci\`encies, Torre C5 parell, 2a pl, 08193 Bellaterra, Spain\\
$^{7}$Astrophysics Group, Keele University, Staffordshire, ST5 5BG, UK\\
$^{8}$SUPA, School of Physics \& Astronomy, University of St. Andrews, North Haugh, KY16 9SS, St. Andrews, Fife, Scotland, UK\\
$^{9}$Cavendish Laboratory, J J Thomson Avenue, Cambridge CB3 0HE, UK\\
$^{10}$University of Bern, Center for Space and Habitability, Sidlerstrasse 5, CH-3012, Bern, Switzerland\\
$^{11}$Department of Physics, University of Warwick, Coventry CV4 7AL, UK\\
$^{12}$N. Copernicus Astronomical Centre, Polish Academy of Sciences, Bartycka 18, 00-716 Warsaw, Poland\\
}
\begin{document}

\date{Accepted ?. Received ?; in original form ?}

\pagerange{\pageref{firstpage}--\pageref{lastpage}} \pubyear{2014}

\maketitle

\label{firstpage}

\begin{abstract}
WASP-80b is a missing link in the study of exo-atmospheres. It falls between the warm Neptunes and the hot Jupiters and is amenable for characterisation, thanks to its host star's properties.
We observed the planet through transit and during occultation with {\it Warm Spitzer}. Combining our mid-infrared transits with optical time series, we find that the planet presents a transmission spectrum indistinguishable from a horizontal line.
In emission, WASP-80b is the intrinsically faintest planet whose dayside flux has been detected in both the 3.6 and 4.5 $\mu$m {\it Spitzer} channels. The depths of the occultations reveal that WASP-80b is as bright and as red as a T4 dwarf, but that its temperature is cooler. {  If planets go through the equivalent of an L-T transition, our results  would imply this happens at cooler temperatures than for brown dwarfs.}
Placing WASP-80b's dayside into a colour-magnitude diagram, it falls exactly at the junction between a blackbody model and the T-dwarf sequence; we cannot discern which of those 
two interpretations is the more likely. WASP-80b's flux density is as low as GJ\,436b at 3.6 $\mu$m; the planet's dayside is also fainter, but bluer than HD\,189733Ab's nightside (in the [3.6] and [4.5] {\it Spitzer} bands).
Flux measurements on other planets with similar equilibrium temperatures are required to establish whether irradiated gas giants, like brown dwarfs, transition between two spectral classes. An eventual detection of methane absorption in transmission would also help lift that degeneracy.

We obtained a second series of high-resolution spectra during transit, using HARPS. We reanalyse the Rossiter-McLaughlin effect.  The data now favour an aligned orbital solution and a stellar rotation nearly three times slower than stellar line broadening implies. A contribution to stellar line broadening, maybe macroturbulence, is likely to have been underestimated for cool stars, whose rotations have therefore been systematically overestimated.
\end{abstract}


\section{Introduction}

Orbital migration, disc-driven or dynamical in origin, has created hot Jupiters \citep{Mayor:1995uq,Lin:1996yq,Rasio:1996ly}. Having a high probability to transit, they opened up the opportunity to study the atmospheric content of extrasolar gas giants without having to spatially resolve them (e.g. \citealt{Seager:2010kx}). In addition, there is hope that the chemical abundances of these planets carries information about where they formed \citep{Oberg:2011qv}. This can be combined with those systems' architectures (multiplicity, eccentricity, orbital inclination) and give a greater appreciation of the processes generating the diversity and width of parameter space that exoplanets occupy.

Their proximity to their host star influences what we can know about these gas giants. For instance, water in the upper atmosphere of the planet will be in the gas phase as a result of the intense irradiation from the host star; its abundance can be measured. This enables us to infer the oxygen content, a quantity which is hard to obtain in the case of a cold planet like Jupiter \citep{Madhusudhan:2014jk}. Then, the carbon to oxygen abundance ratio can be quantified and linked to the chemistry of the protoplanetary disc and ultimately, to the formation process of the planet \citep{Bolton:2010xy}.

Proximity also fosters the appearance of a host of exotic phenomena which remain to be fully investigated. For instance, irradiation causes hot Jupiters to be inflated, whose exact cause (or causes) continues to be an active area of research (e.g. \citealt{Demory:2011lr} and references there-in). The temperatures of the majority of hot Jupiters whose atmospheres have been studied, are in the range 1,500 to 2,000 K. Their densities are lower than colder planets which makes them practical targets for transmission spectroscopy, at transit. Their bloated size and elevated temperature are also advantages when carrying emission spectroscopy, at occultation.

Hot Jupiters have sizes and dayside temperatures that makes them resemble late M-dwarfs and L-type brown dwarfs. They share a common colour-magnitude space \citep{Triaud:2014kq}. These non-irradiated, self-luminous objects are covered by dust clouds, produced by the condensation of various species {  such as silicates, aluminium and iron bearing molecules as well as }titanium and vanadium oxides \citep{Kirkpatrick:2005th}, which become patchy with decreasing temperatures \citep{Ackerman:2001fk,Burgasser:2002kx,Artigau:2009lr,Radigan:2012rt,Gillon:2013qv,Buenzli:2014fj,Crossfield:2014db}. 
{  In the near-infrared bands $Y$, $J$, $H$, and $K$, the L-T transition is characterised by a blueward shift where the hotter, deeper parts of the atmosphere start becoming visible through cloud clearings. In the mid-IR [3.6] and [4.5 $\mu$m] bands, that transition is defined by a redward slope, produced in part by a shift of the blackbody peak towards cooler temperatures but also, by an increase in CH$_4$ absorption (and its associated reduction in CO and CO$_2$) \citep{Kirkpatrick:2005th}. }

The presence of clouds and hazes is suspected to hamper the detection of molecules in the transmission spectra of hot Jupiters  (e.g. \citealt{Sing:2011ly,Jordan:2013dn,Mandell:2013fk,Pont:2013lr}). If the analogy between self luminous ultra-cool dwarfs and irradiated gas giants holds, then planets with daysides similar to mid to late T-dwarfs ought to be mostly cloud free, presenting to our line of sight a clear terminator with strong methane absorption. {  In the two {\it Spitzer} bands we used,} a T-dwarf like planet would have an illuminated face redder than the typical hot Jupiter when using  the two {\it Warm Spitzer} channels \citep{Triaud:2014ly}.

WASP-80b is a special planet. This gas giant, discovered by the WASP collaboration \citep{Pollacco:2006fj,Collier-Cameron:2007fj,Hellier:2012fj}, orbits a bright (V = 11.9, K = 8.4, W1 = 8.28, W2 = 8.31) late K dwarf/early M-dwarf. Despite a 3~day orbit, typical of hot Jupiters, it has an equilibrium temperature close to only 800 K \citep{Triaud:2013sk}. This is one of the coolest gas giants whose atmosphere can be studied in transmission and emission using current instrumentation. As such, it is an obvious target to observe to verify how temperature affects planetary spectra and to continue the exploration in colour-magnitude space. This is one of only a few systems expected to resemble a T-dwarf \citep{Triaud:2014ly}. In addition, of the other planets with similar or colder temperatures that have been studied, all are Neptune-like planets, orbiting M-dwarfs. WASP-80b fills an empty space between those small, warm Neptunes, and the more massive and large, hot Jupiters. The warm Neptunes {  mostly} have featureless transmission spectra \citep{Kreidberg:2014xy,Knutson:2014fk,Ehrenreich:2014jk}{ , with the notable exception of HAT-P-11b \citep{Fraine:2014lr}}. WASP-80b has a larger {  scale-height}, more amenable for transmission spectroscopy.

To improve the characterisation of this fascinating planet, we  observed some of its transits and occultations at 3.6 and 4.5 $\mu$m with {\it Spitzer}, and gathered new ground-based transit photometry and spectroscopy. We combine these new observations with published high-quality data, and perform a global analysis of the resulting extensive data set. We present here the results of this work. Our paper is simply organised: we first present our data collection and analysis, and then discuss our results and conclude.

\begin{figure*}  
\begin{center}  
	\begin{subfigure}[b]{0.49\textwidth}
		\caption{bin = 2.5 min}
		\includegraphics[width=\textwidth]{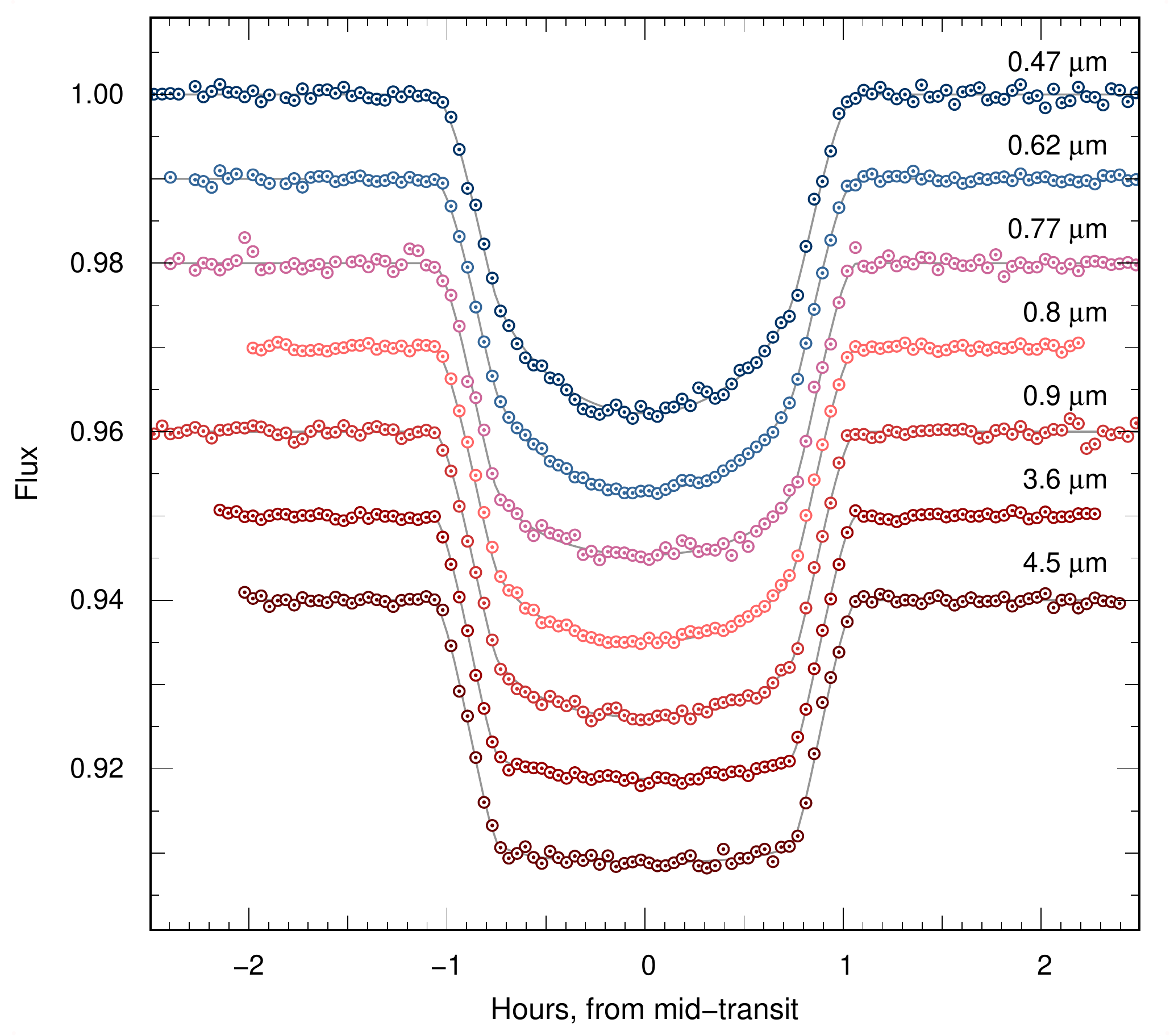}  
		\label{fig:trans}  
	\end{subfigure}
	\begin{subfigure}[b]{0.49\textwidth}
		\caption{bin = 5 min}
		\includegraphics[width=\textwidth]{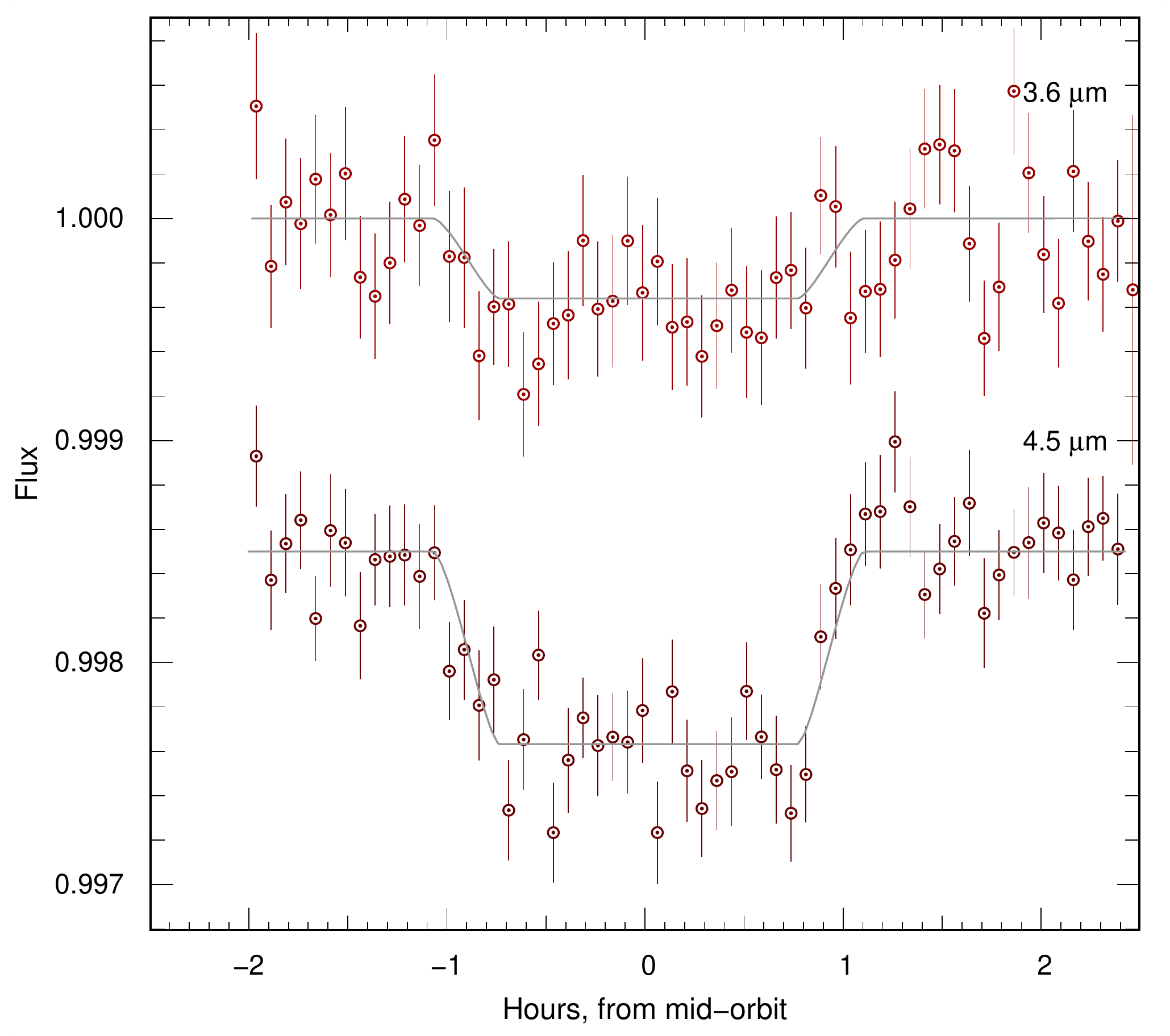}  
		\label{fig:occc}  
	\end{subfigure}
	\caption{Photometric timeseries covering the transit (left panel) and occultation (right panel) of WASP-80. For visual convenience, the data are presented binned and vertically offset. They are labeled by central wavelength of the broad-band filter used. Un-corrected and un-binned data are available in the appendices.
}\label{fig:data}  
\end{center}  
\end{figure*}

\section{Data collection and analysis}

We used the photometric and spectroscopic data presented in \citet{Triaud:2013sk} as well as the lightcurves collected and analysed by \citet{Mancini:2014lr} (see Table~\ref{tab:phot} and Fig.~\ref{fig:ground}). To these, we add ground and space-based photometry, described in the first subsection, and new spectroscopic information reported in the second subsection.

\subsection{Broad band photometry at transit and at occultation}\label{sec:phot}

\subsubsection{New Sloan-{\it z}'  and Gunn-{\it r}' transit photometry with TRAPPIST and {\it Euler}}

We captured three additional transits during the 2013 season, using TRAPPIST \citep{Jehin:2011dk}. One of these -- on the night starting on 2013-06-15 -- was acquired in tandem with the {\it Euler} telescope. They are also simultaneous with the DFOSC and GROND lightcurves presented by \citet{Mancini:2014lr}, and with the Rossiter-McLaughlin effect described in the following subsection. That night, five telescopes observed WASP-80b simultaneously from La Silla.

The TRAPPIST observations made use of a Sloan $z$' filter, with each frame exposed for 13s.
The transit observed with {\it Euler} was acquired through a Gunn-$r$' filter, and an exposure time of 50s.
Observations were set up similarly on both telescopes. Notably, we kept the positions of each star on the CCD chip within a box of a few 
pixels. This is achieved using a `software  guiding' system which regularly derives  an astrometric 
solution on the science images, and sends pointing corrections to the mount when needed. This improves the photometric precision by reducing the instrumental noise due to flat-fielding errors. Table~\ref{tab:phot} presents a log of the observations. 

After a standard pre-reduction (bias, dark, flat-field correction), stellar fluxes were extracted from our 
images using the {\tt IRAF/DAOPHOT}\footnote{{\tt IRAF} is distributed by the National Optical Astronomy 
Observatory, which is operated by the Association of Universities for Research in Astronomy, Inc., under 
cooperative agreement with the National Science Foundation.} aperture photometry software \citep{Stetson:1987kl}. 
For each transit, several sets of reduction parameters were tested. We kept the set yielding the most precise 
photometry on the stars having a brightness similar to WASP-80. After a careful selection of reference stars, differential 
photometry was then obtained. The resulting light curves are shown in the appendix (Fig.~\ref{fig:ground}).

\subsubsection{{\it Spitzer} transit and occultation photometry at 3.6 and 4.5 $\mu$m}

We requested time to observe WASP-80 with {\it Spitzer} (PID 90159; Triaud) and captured 
six eclipses of the system with IRAC \citep{Fazio:2004fk}. We obtained data covering one transit and two occultations at 3.6 $\mu$m and 
one transit and two occultations at 4.5 $\mu$m. The scheduling of these observations was based on the orbital solution 
presented by \citet{Triaud:2013sk}. We used the subarray mode ($32\times32$ pixels, 1.2 arcsec/pixel) with exposure times of 0.4s (3.6 $\mu$m) and 2s (4.5 $\mu$m). The resulting sets of 64 individual subarray images were calibrated by the {\it Spitzer} pipeline version S19.1.0 and are available on the {\it Spitzer} Heritage Archive Database\footnote{\href{http://sha.ipac.caltech.edu/applications/Spitzer/SHA}{http://sha.ipac.caltech.edu/applications/Spitzer/SHA}}.

The data were reduced in the same way for all six time series. We first converted fluxes from the {\it Spitzer} units of specific
intensity (MJy sr$^{-1}$) to photon counts. We then performed aperture photometry on each subarray image with
{\tt IRAF/DAOPHOT}. We tested different
aperture radii and background annuli, and found the best results
with an aperture radius of 2.5 pixels and a background annulus
extending from 11 to 15.5 pixels from the centre of the point-spread
function (PSF). We measured the centre and width of the PSF (FWHM)
by fitting a 2D-Gaussian profile on each image. We then inspected
at the $x$-$y$ distribution of the measurements, and discarded the
few measurements having a visually discrepant position relative
to the bulk of the data. For each block of 64 subarray images,
we then discarded the discrepant values for the measurements of
flux, background, $x$ and $y$ positions, and PSF widths in the $x$- and
$y$-direction, using a 10-$\sigma$ median clipping for the six parameters.
We averaged the remaining values, taking the errors on the average
flux measurements as photometric errors. At this stage, we
used a moving median filter in flux on the resulting light curves
to discard outlier measurements due to cosmic hits, for example.
The resulting lightcurves can be inspected in the appendix, in Figs.~\ref{fig:raw_trans} and \ref{fig:raw_occ}.

\subsection{HARPS observations at transit}\label{sec:ros}

On the night starting on 2013-06-15, simultaneously with much of the ground-based photometry, we obtained a series of 26 spectra with HARPS (Prog.ID 091.C-0184; Triaud) at a cadence of approximately 10 minutes. These new data are of higher quality than the transit spectroscopy presented in \citet{Triaud:2013sk} as a result of better weather conditions (clear sky, 0.6 arcsecond seeing) and higher cadence. Seven additional spectra were obtained on the nights leading to and following the transit night. They helped further contain the orbital parameters. 

The data were reduced using the standard HARPS reduction software, and the radial velocities were extracted by cross-correlating each spectrum with a K5 mask, as was done in \citet{Triaud:2013sk}.

\begin{table}[h] 
  \caption{Results of our MCMC fits to the photometric and spectroscopic data for WASP-80. Errors on the last two digits of each parameters, are given in brackets. Asterisks mark
   parameters which are controlled by priors.
   }\label{tab:res}
  \begin{tabular}{llrl}
  \hline
  \hline
  Parameters & Units & Values &\\
  \hline
  {\it the star}\\
$T_{\rm eff}$		&K				&$4143_{(-94)}^{(+92)}$	&*\\
$\rm[Fe/H]$		&dex				&$-0.13_{(-17)}^{(+15)}$	&*\\
$\log g_\star$		&cgs				&$4.663_{(-16)}^{(+15)}$	\\
$M_\star$			&$M_\odot$		&$0.577_{(-54)}^{(+51)}$	& *\\
$R_\star$			&$R_\odot$		&$0.586_{(-18)}^{(+17)}$	\\
$\rho_\star$		&$\rho_\odot$		&$2.875_{(-86)}^{(+55)}$		\\
$ v \sin i_\star$		&km s$^{-1}$		&$1.27_{(-17)}^{(+14)}$	\\
\hline
{\it the planet}\\
$ P$				&day				&$3.06785234_{(-79)}^{(+83)}$	\\
$ T_0$			&BJD			&$6487.425006_{(-25)}^{(+23)}$	\\
$ D$	(3.6 $\mu$m)	&--				&$0.02937_{(-13)}^{(+13)}$	\\
$ W$				&day				&$0.08878_{(-14)}^{(+13)}$	\\
$ b$				&$R_\odot$		&$0.215_{(-22)}^{(+20)}$	\\
$ K$				&m s$^{-1}$		&$109.0_{(-4.4)}^{(+3.1)}$	\\
\\
$\log g_{\rm p}$	&cgs				&$3.145_{(-16)}^{(+15)}$	\\
$ a/R_\star$		&--				&$12.63_{(-13)}^{(+08)}$	\\
$M_{\rm p}$		&$M_{\rm Jup}$	&$0.538_{(-36)}^{(+35)}$	\\
$R_{\rm p}$ (3.6 $\mu$m)	&$R_{\rm Jup}$	&$0.999_{(-31)}^{(+30)}$	\\
$\rho_{\rm p}$		&$\rho_{\rm Jup}$	&$0.539_{(-24)}^{(+29)}$	\\
$T_{\rm eq}$		&K				&$825_{(-19)}^{(+19)}$	\\
\\
$ a$				&AU				&$0.0344_{(-11)}^{(+10)}$	\\
$ i_{\rm p}$		&deg			&$89.02_{(-10)}^{(+11)}$	\\
$ \beta$			&deg			&$14_{(-14)}^{(+15)}$	\\
$ e$				&--				&$0.002_{(-02)}^{(+10)}$	\\
$ \omega$		&deg			&$94_{(-21)}^{(+120)}$	\\
\hline
\multicolumn{3}{l}{\it star to planet ratio, $R_{\rm p} / R_\star$}\\
0.47 $\mu$m		&--			&$0.17006_{(-78)}^{(+77)}$	\\
0.62 $\mu$m		&--			&$0.17169_{(-62)}^{(+63)}$	\\
0.77 $\mu$m		&--			&$0.17220_{(-67)}^{(+64)}$	\\
0.80 $\mu$m		&--			&$0.17173_{(-56)}^{(+56)}$	\\
0.90 $\mu$m		&--			&$0.17127_{(-69)}^{(+69)}$	\\
3.6 $\mu$m		&--			&$0.17137_{(-39)}^{(+37)}$	\\
4.5 $\mu$m		&--			&$0.17230_{(-39)}^{(+40)}$	\\
\hline
\multicolumn{3}{l}{\it occultation depths}\\
3.6 $\mu$m		&ppm			&$455_{(-100)}^{(+100)}$	\\
4.5 $\mu$m		&ppm			&$944_{(-65)}^{(+64)}$	\\
\hline
\multicolumn{3}{l}{\it brightness temperatures}\\
3.6 $\mu$m		&K				&$901_{(-72)}^{(+68)}$	\\
4.5 $\mu$m		&K				&$888_{(-57)}^{(+58)}$	\\
\hline
\end{tabular}
\end{table}

\subsection{Global data analysis}\label{sec:analysis}

Strong constraints on the system parameters can be derived by 
performing a global Bayesian analysis of the photometric and radial velocity time series. We constructed the posterior probability distributions functions of the global model 
parameters using an adaptive Markov Chain Monte-Carlo (MCMC) code described 
in \citet{Gillon:2014ix} (and references therein). 

Transits and occultations were modelled with the algorithm written by \citet{Mandel:2002kx}. Each light curve is multiplied by a baseline model, which accounts for 
other astrophysical and instrumental effects resulting in  
photometric variations. A quadratic limb-darkening law was assumed for the transits.
For each light curve, a baseline model (see Table~\ref{tab:phot}) was selected by way of 
minimising the Bayesian Information Criterion (BIC; \citealt{Schwarz:1978zz}).
The radial velocities were modelled by a Keplerian orbit for the star combined
to a systemic velocity, and to the spectroscopic transit model of \citet{Gimenez:2006kx}.

For the ground-based light curves, a second-order time polynomial is systemically 
assumed for each baseline. It accounts for a possible low-frequency stellar variability, for a difference in colour between the target and its comparison stars, and for their
consequent differential extinction. The modelling of other effects (dependance of 
fluxes on FWHM, on position, meridian flip) were also required for a few light curves 
(see Table~\ref{tab:phot}). 

For the {\it Spitzer} photometry, our baseline models representing the well-documented
`ramp' and `phase-pixel' effects (e.g \citealt{Knutson:2008qy,Lewis:2013qy}), are similar to those employed 
in the analysis of GJ\,1214  \citep{Gillon:2014ix}; we refer the reader to that paper for a 
detailed description.

The only informative prior probability distribution functions assumed in our analysis are 
for the stellar mass $M_\star$, effective temperature $T_{\rm eff}$,  metallicity ${\rm [Fe/H]}$, and for the
limb-darkening coefficients of each bandpass. We
assumed normal prior distributions $N(0.58,0.05^2)$ $M_\odot$, $N(4145, 100^2)$ K, and $(-0.14,0.16)$ dex, for $M_\star$, $T_{\rm eff}$, and ${\rm [Fe/H]}$  respectively \citep{Triaud:2013sk}. Normal prior distributions are also assumed for the coefficients $u_1$ and $u_2$ of the quadratic limb darkening law. The parameters of our Gaussians
were interpolated from the tables of \citet{Claret:2011rm}  for each of the
bandpasses we observed in, and corresponding to a star with $T_{\rm eff}=4145\pm100$K, $\log{g_\star} = 4.689\pm0.013$ dex, and 
${\rm [Fe/H]}=-0.14\pm0.16$ dex \citep{Triaud:2013sk}. No priors were applied to the Rossiter-McLaughlin effect\footnote{The Rossiter-McLaughlin effect was analysed more in detail separately from the global analysis. The description is provided in Sec.~\ref{sec:ros}.}.

A first chain of 10\,000 steps was launched. It brought to our attention the need to rescale our photometric 
errors in order to account for an extra-white noise and for some correlated noise. We also estimated the amount of stellar noise (``jitter'') in the radial velocities and added this quadratically to the radial velocity errors (see \citealt{Gillon:2012fj} for details). The error rescaling factors for the 
light curves are given in Table~\ref{tab:phot}. For the CORALIE radial velocities, the deduced jitter noise was 8.1 m s$^{-1}$; the HARPS data required none. With errors corrected, 
we performed our main analysis that consisted of two chains composed of 100\,000 steps. They converged
according to the statistical test of \citet{Gelman:1992rt}. Table~\ref{tab:res} collects our derived parameters and
their $1\sigma$ confidence region. 

The resulting light curves, corrected from systematic effects and and co-added according to wavelength, are displayed in Fig.~\ref{fig:data}. The transits (Fig.~\ref{fig:trans}) have been binned in 2.5 minute bins (for visual convenience only), and the occultations (Fig.~\ref{fig:occc}) in 5 minute bins.

\subsection{A study of the Rossiter-McLaughlin effect}\label{sec:ros}

In the discovery paper, we were unable to distinguish the impact parameter from zero, which led the fitting procedure of the Rossiter-McLaughlin effect to produce degenerate solutions for the projected spin--orbit angle $\beta$ (e.g. \citealt{Triaud:2011vn,Albrecht:2011rt}). Here we study several possibilities, getting further into some details of the Rossiter-McLaughlin effect that the global analysis did not touch.

The Rossiter-McLaughlin effect was modelled using the prescription of \citet{Gimenez:2006kx}, in the same manner as in \citet{Lopez-Morales:2014qv}, using priors as input for all the parameters except those controlling the effect. Both time series were adjusted as a single dataset, since no offset was detected between them. As in \citet{Triaud:2013sk}, we investigated three different propositions: 
\begin{itemize}
\item using no priors on the projected stellar rotation $v \sin i_\star$; 
\item using a prior, obtained from spectral line broadening, $v \sin i_\star = 3.55 \pm 0.33$ km s$^{-1}$ \citep{Triaud:2013sk}; 
\item applying a prior on $\beta = 0 \pm 10^\circ$.
\end{itemize}

The first and third propositions produced similar, aligned, solutions, whereas the second converged on a inclined orbit. Thanks notably to the {\it Spitzer} photometry, which estimated a non-zero impact parameter,  $\beta$ could be resolved ($54\pm7^\circ$). However, this solution yields a poorer fit ($\Delta \chi^2 = 30.7$, for the same number of degrees of freedom) in part because it remains in $2\sigma$ tension with the adopted prior, supporting incompatibility between the value measured thanks to the Rossiter-McLaughlin effect and that estimated from stellar line broadening. We thus favour the results from the first proposition ($\beta = 14\pm14^\circ$, $v \sin i_\star = 1.27 \pm 0.14$ km s$^{-1}$) and adopt them; they can be found in Table \ref{tab:res}. Our estimation of $v \sin i_\star$ is thus an independent measurement.

\begin{figure}
\center
\includegraphics[width= \linewidth]{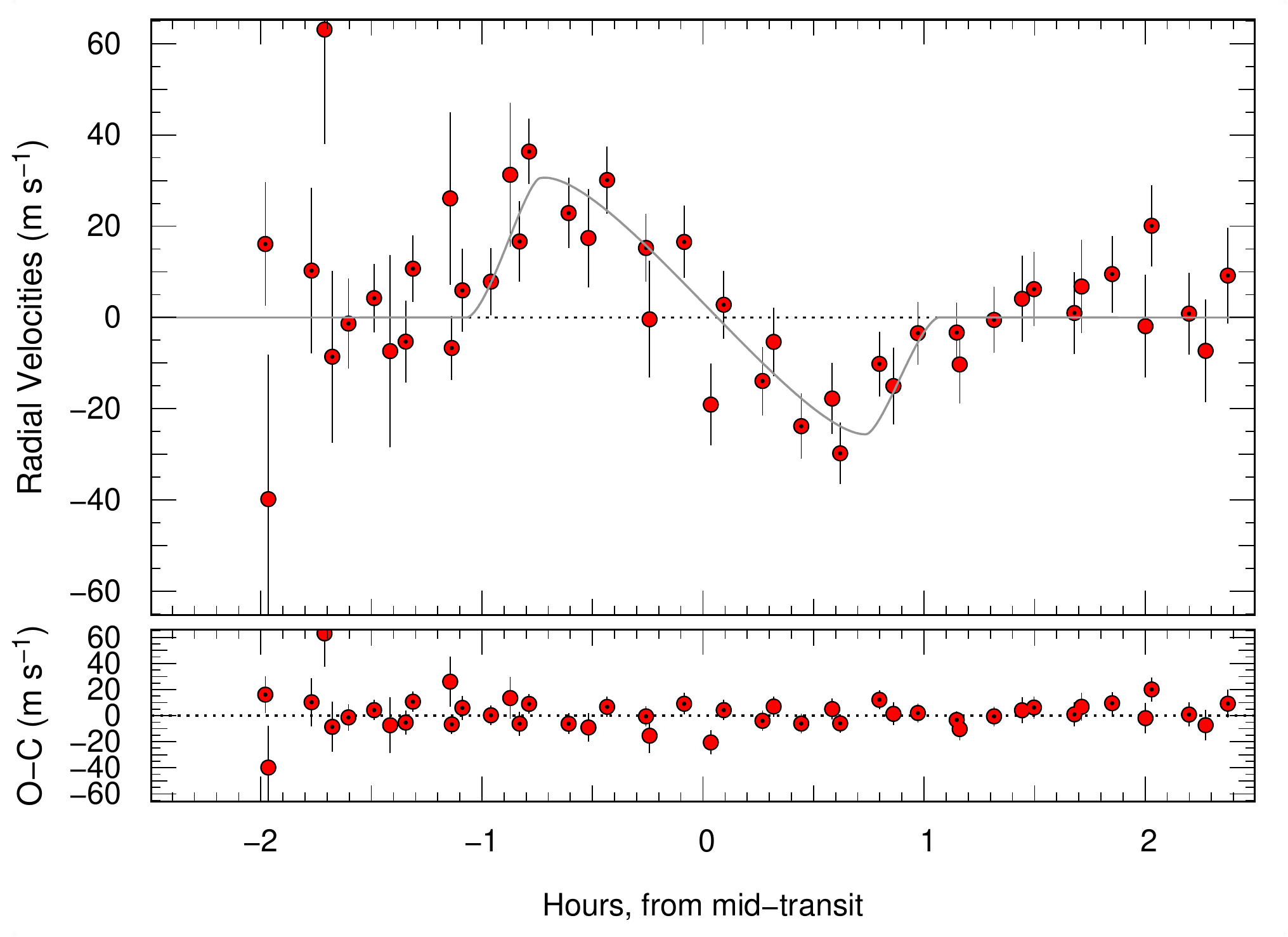}
\caption{The Rossiter-McLaughlin effect caused by WASP-80b, as observed by HARPS, its most likely model and its corresponding residuals. The radial velocities are corrected for the Doppler reflex motion. Additional dark dots in the centre of some data points indicate the new data.
}
\label{fig:ros}
\end{figure}

The most likely fit and the data are displayed in Fig.~\ref{fig:ros}. 

\begin{figure}
\center
\includegraphics[width= \linewidth]{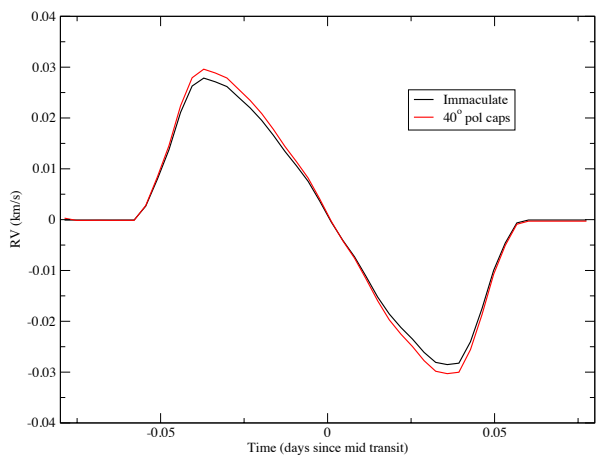}
\caption{Simulation of the Rossiter-McLaughlin effect on a star similar to WASP-80. Two configurations are tested here. One where the star has no spots, and one where it possesses spots at the poles covering the upper and lower 40 degrees in latitude. This increases slightly the amplitude of the effect, because the fractional contribution to the stellar rotation hidden by the planet has gained in importance.
}
\label{fig:ros_simu}
\end{figure}

\section{Discussion}

\subsection{About the Rossiter-McLaughlin effect}

Thanks to the quality of our photometric data at transit and the much reduced influence of limb darkening in the mid-infrared, it was possible to precisely measure the impact parameter of WASP-80b. This lifts the degeneracy between $v \sin i_\star$ and $\beta$ noted in \citet{Triaud:2013sk}, and enables us to show that the planet is on a coplanar orbit instead of inclined, as we had previously reported. The amplitude of the Rossiter-McLaughlin effect implies a slower rotation for the star than is estimated from stellar line broadening. The values are incompatible at the 7$\sigma$ level. In  \citet{Triaud:2013sk}, we took this difference as a sign that the planet's orbital plane was severely inclined (which the poor precision on the impact parameter then allowed), the data are now of sufficient quality to instead favour a coplanar solution. This brings this system more in line with the patterns of spin--orbit alignement with stellar host parameters proposed by  \citet{Schlaufman:2010fk} and \citet{Winn:2010rr}.

We attempted to understand this discrepancy by verifying how $v \sin i_\star$ from spectral line broadening is affected by activity, and what the resulting Rossiter-McLaughlin effect would look like. To do this, we used tools developed in \citet{Herrero:2014lr}. 

The star is chromospherically active \citep{Triaud:2013sk,Mancini:2014lr}, and we could also detect, from the CORALIE spectra, that the $H_{\alpha}$ equivalent width of WASP-80 has a wider variation epoch to epoch than intra-night measurements show, but presents no periodicity. Despite these indications of stellar activity, there is no rotational modulation in the photometry \citep{Triaud:2013sk}.  Some spot configurations can lead to weak photometric signal: 1) well centred and stable polar spots, 2) a continuously spotted active latitude, or 3) a near homogeneously spotted surface. Those configurations were simulated, and produced slight differences in the estimation of $v \sin i_\star$, but the differences are less than 100 m s$^{-1}$. Fig. \ref{fig:ros_simu} shows a simulated Rossiter-McLaughlin effect for a spotless star, and one with large polar spots. The difference between both models is of order 1 m s$^{-1}$ -- well within error bars. 

{  Furthermore, stellar spots outside of the transit chord produce a slope in the transmission spectrum \citep{Jordan:2013dn,McCullough:2014lr}. The important  coverage necessary to significantly alter the Rossiter-McLaughlin effect would presumably leave a trace that is not detected in the multi-wavelength photometry presented in Fig.~\ref{fig:transmi}}.

\begin{figure*}  
\begin{center}  
	\begin{subfigure}[b]{0.45\textwidth}
		\includegraphics[width=\textwidth]{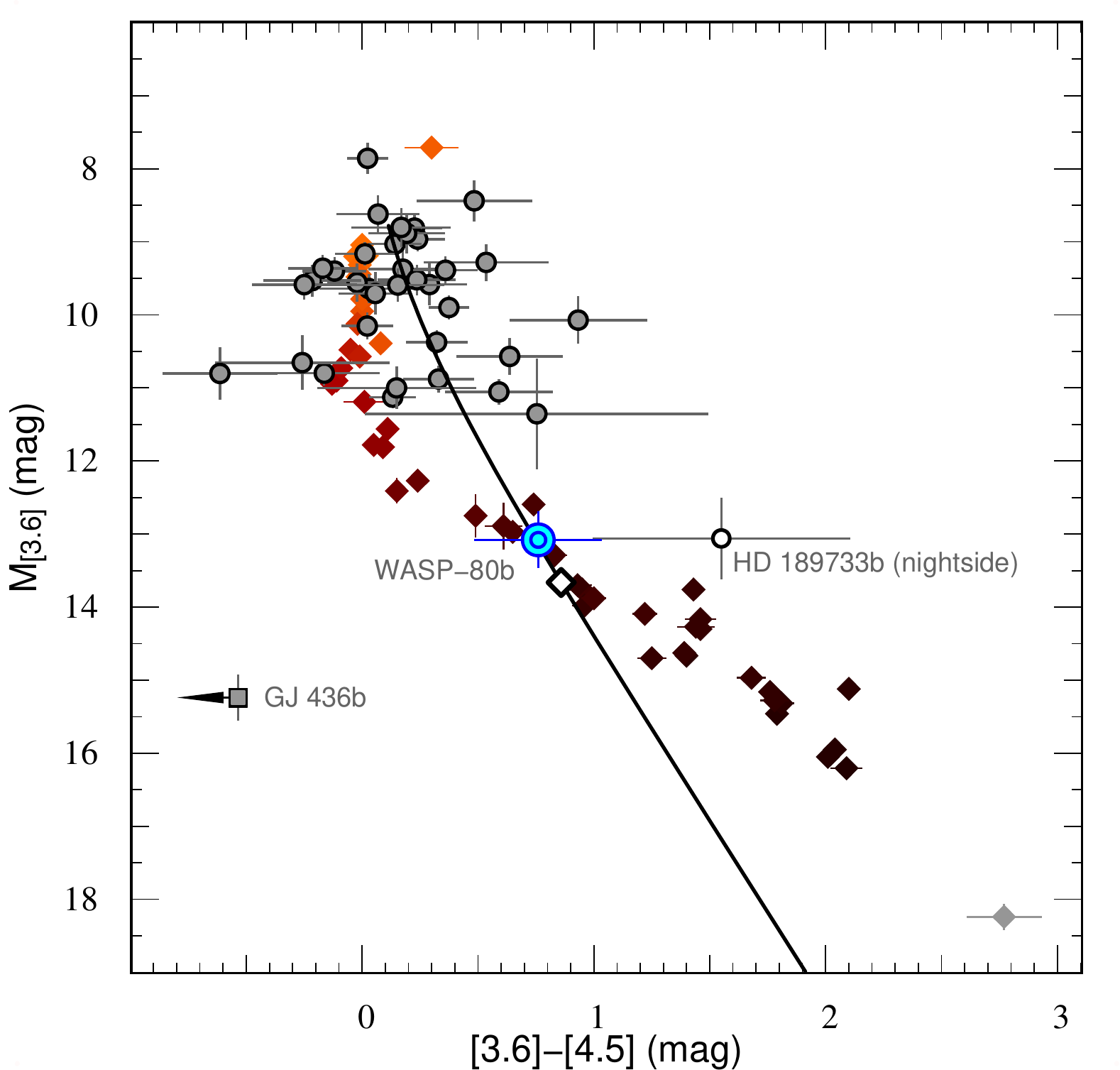}  
		\label{fig:CM1}  
	\end{subfigure}
	\begin{subfigure}[b]{0.45\textwidth}
		\includegraphics[width=\textwidth]{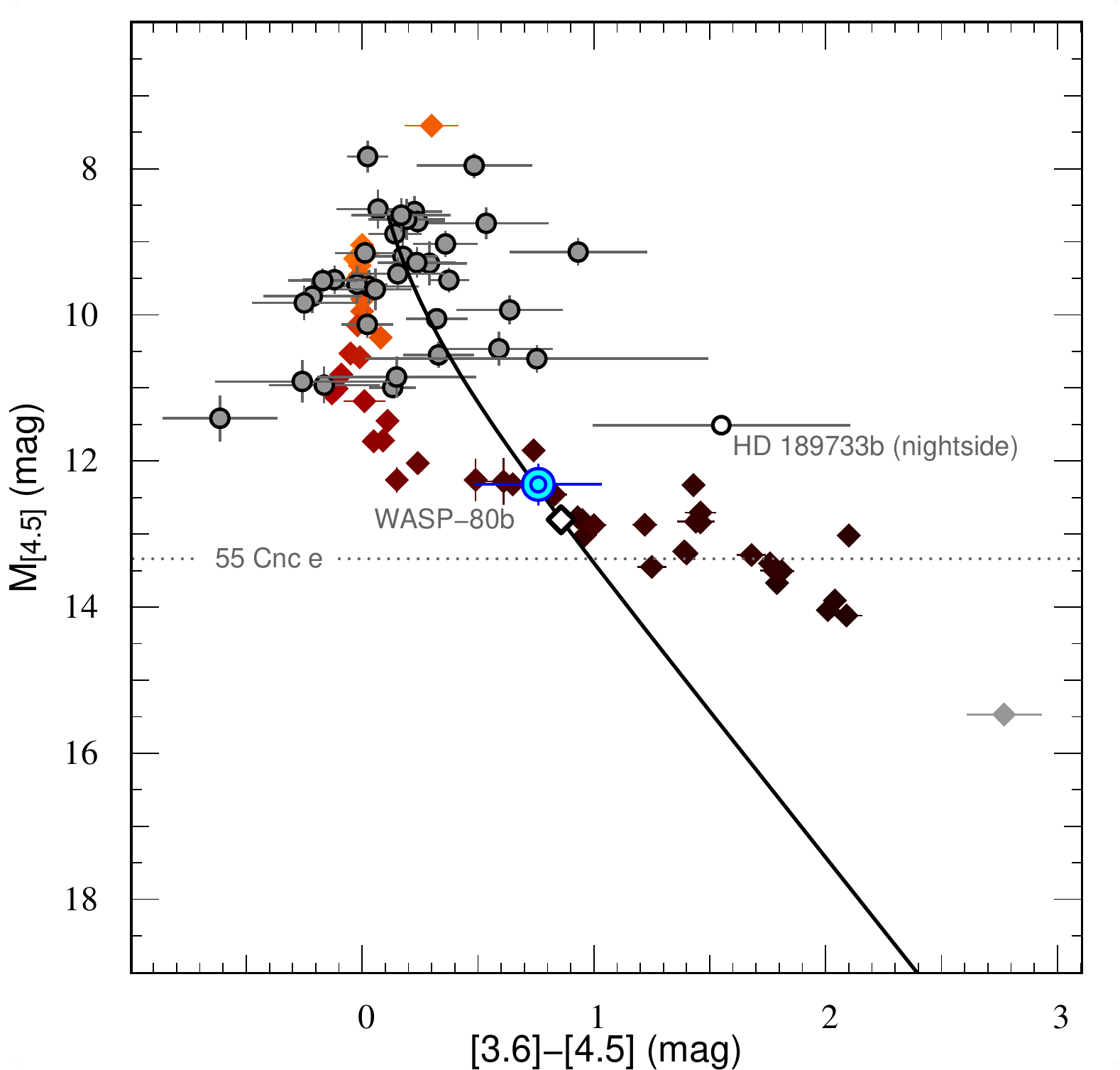}  
		\label{fig:CM2}  
	\end{subfigure}
	\begin{subfigure}[b]{0.9\textwidth}
		\includegraphics[width=\textwidth]{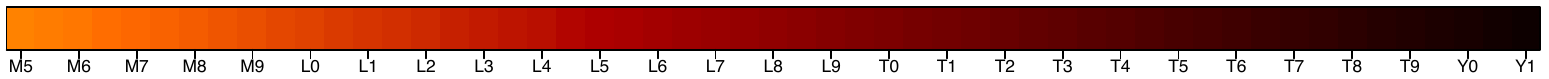}  
		\label{fig:bar}  
	\end{subfigure}	
	\caption{Two colour-magnitude diagrams highlighting the position of the dayside of WASP-80b. Circles represent planets; GJ 436b and the nightside of HD\,189733Ab have slightly different symbols to attract attention. Diamonds show the location of ultra-cool dwarfs obtained from \citet{Dupuy:2012lr}, colour-coded as a function of spectral type. Grey symbols indicate no spectral classification. The blackline outlines the location of blackbodies with of a size similar to WASP-80b, but with effective temperatures ranging from 4,000K to 400K. The empty diamond on this line show 800\,K. WASP-80b has one of the reddest dayside amongst all hot Jupiters, but remains bluer than HD\,189733Ab's nightside.
}\label{fig:CM}  
\end{center}  
\end{figure*}

Resolving this $v\,\sin\,i_\star$ discrepancy is outside of the range of topics this paper aimed to study, however we note that most laws attempting to replicate macroturbulence in stars suggest this phenomenon's contribution to stellar line broadening decreases with stellar effective temperature, and vanishes for the coolest stars (e.g. \citealt{Gray:2008fj}). \citet{Doyle:2014qf} have instead shown that it may plateau to a set value. Sadly, WASP-80 is too cool and falls out of their revised calibration. Our issues only highlights how tricky it is to estimate $v \sin i_\star$ from stellar line broadening. {  It  reminds us of the vigilance} we need to keep when including these values as priors in an analysis of the Rossiter-McLaughlin effect. Still, the fact remains that WASP-80 has wide lines compared to others stars with a similar $B-V$ \citep{Triaud:2013sk}. If its width is equivalent to $v \sin i_\star = 1.3$ km s$^{-1}$ then all other late K and early M-dwarfs must be rotating much slower than they have been thought to do until now. As such, the Rossiter-McLaughlin effect may become a means to study the effect of macroturbulence by providing an independent measurement of $v \sin i_\star$.

\begin{table}
  \caption{Predicted magnitudes and occultation depth for various broad bands, and for two scenarios representing the two alternatives for WASP-80b: a T4 spectrum, or a blackbody.
  To compute the magnitude for the T4 type, we used the relations produced in \citet{Dupuy:2012lr}, with their errors. The uncertainty on the distance is propagated to provide a credible range of values. Our detections fall within those boundaries. Those errors are only useful when considering each band in isolation.  
  Occultations depth were estimated using stellar visual magnitudes obtained from WISE \citep{Cutri:2013fj} and 2MASS \citep{Cutri:2003rt}.
   }\label{tab:prediction}
  \begin{tabular}{lrrr}
  \hline
  \hline
   & absolute  & visual  & occultation \\
  band &  magnitude &  magnitude &  depth [ppm]\\
  \hline
&\multicolumn{3}{l}{\it if a T4}\\
$J_{\rm 2MASS}$		&$14.3 \pm 0.4$	&$17.9\pm0.5$		&$340\pm180$\\
$H_{\rm 2MASS}$		&$14.1 \pm 0.4$	&$17.7\pm0.5$		&$210\pm120$\\
$K_{\rm 2MASS}$		&$14.1 \pm 0.4$	&$17.7\pm0.5$		&$182\pm98$\\
$[3.6]$				&$12.9 \pm 0.3$	&$16.5\pm0.4$		&$520\pm210$\\
$[4.5]$				&$12.3 \pm 0.2$	&$15.9\pm0.3$		&$930\pm270$\\
$[5.8]$				&$12.0 \pm 0.3$	&$15.6\pm0.4$		&$1150\pm470$\\
  \hline
&\multicolumn{3}{l}{\it if a 900K blackbody}\\
$J_{\rm 2MASS}$		&$20.7$		&$24.3\pm0.25$		&$1.0\pm0.2$\\
$H_{\rm 2MASS}$		&$17.5$		&$21.1\pm0.25$		&$9.2\pm2.2$\\
$K_{\rm 2MASS}$		&$15.5$		&$19.1\pm0.25$		&$50\pm12$\\
$[3.6]$				&$13.0$		&$16.6\pm0.25$		&$480\pm110$\\
$[4.5]$				&$12.3$		&$15.9\pm0.25$		&$930\pm220$\\
$[5.8]$				&$11.7$		&$15.3\pm0.26$		&$1510\pm370$\\
\hline
\end{tabular}
\end{table}

\subsection{About the colour of WASP-80b}

Flux-drops during occultation can be combined with the apparent magnitudes of WASP-80, as measured by {\it WISE} \citep{Wright:2010qv} in its W1 \& W2 channels (equivalent to {\it Spitzer}'s channels 1 \& 2; \citealt{Triaud:2014ly}) to compute WASP-80b's dayside apparent magnitudes: $m_{[3.6]} = 16.63\pm0.26$ and $m_{[4.5]} = 15.872\pm0.078$. This implies a very red colour for WASP-80b: [3.6]-[4.5] = $0.76\pm0.27$, compared to the average colour of hot Jupiters of  $0.149\pm0.025$ (sample taken from \citet{Triaud:2014ly}, except for one significantly updated value on HD\,209458b; \citealt{Zellem:2014rr}). 

We reproduced the methodology applied in \citet{Triaud:2014ly} and find that the WASP-80 system has a distance modulus $m-M = 3.55\pm0.26$ ($51.3\pm6.2$ pc). The absolute magnitudes of WASP-80b's dayside are therefore: $M_{[3.6]} = 13.08\pm0.37$ and $M_{[4.5]} = 12.32\pm0.27$. This is slightly fainter than the nightside of HD\,189733Ab; it is also bluer in those specific bands \citep{Knutson:2012ys,Triaud:2014kq}. 

{  The {\it Spitzer} occultations correspond to brightness temperatures of $901_{-72}^{+68}$ and $888_{-58}^{+57}$ K for channel 1 and 2 respectively, slightly above its equilibrium temperature of 800\,K. These values have been computed using the stellar models, computed by \citet{Kurucz:1993qy}, that were compatible with the stellar parameters that we have used for WASP-80. The error budget contains the observational uncertainties on the occultations' depths and on the stellar parameters, as in \citet{Demory:2013uq}. }

\citet{Dupuy:2012lr} present relations between absolute magnitudes in certain bandpasses, including those we observed in, and spectral classification of ultra-cool dwarfs. The dayside of WASP-80b corresponds to a T4 dwarf, in both the [3.6] and [4.5] channels. {  T4 is the latest type considered within the L-T transition.  The T4 type comprises effective temperatures ranging between 1100 and 1300\,K  \citep{Stephens:2009lr}. 
A 200\,K increase can be compensated by a 30\% reduction in radius and produce a similar luminosity. According to \citet{Baraffe:2003gf}, brown dwarfs can be 0.75~$R_{\rm Jup}$, which would make WASP-80b's dayside fit with a T4's flux, within uncertainties. Provided that this is indeed the case, our results may imply that for planets, the L-T transition happens at cooler temperatures than for brown dwarfs\footnote{{  that transition is expected to occur around 500K \citep{Zahnle:2014fk}. However their work was carried for self-luminous objects only.}}.
}

We do not intend here to attach a spectral classification to WASP-80b, but only to provide a reference point. These relations have been built from --and are therefore only valid for-- ultra-cool dwarfs: self-luminous objects, whose stronger gravities impact the structure and type of clouds layers. {  While for brown dwarfs, a change in composition leads to small variations in the emerging spectrum, planetary spectra can be severely affected. This is particularly relevant since planets cover an important range in metal enrichment  as seen in the Solar system.} For similar composition, irradiation changes a planet's spectrum dramatically too \citep{Burrows:2014yu} and different chemistry can be expected between the night and day sides of a planet, rendering the classification of a planet arduous at best \citep{Cooper:2006yu}. Indeed, the phase curve  of HD\,189733Ab presents evidence that its atmosphere goes through a change in colour similar to a spatial transition between L to T \citep{Triaud:2014kq}.

Our values were placed into two colour-magnitude diagrams (Fig. \ref{fig:CM}), presenting the same data as in \citet{Triaud:2014ly} (except for HD\,209458b; \citealt{Zellem:2014rr} and the nightside of HD\,189733Ab; \citealt{Triaud:2014kq}). WASP-80b sits just on the T-dwarf sequence, which confirms our earlier statement about its resemblance to a T4 spectral type. The redward direction of the T-dwarf sequence (in the [3.6]-[4.5] colour) is characterised by the appearance of strengthening CH$_4$ absorption in the 3.6 $\mu$m channel ({  and a corresponding decrease in CO:} see \citealt{Patten:2006uq,Zahnle:2014fk}). 

Two blackbody loci, corresponding to a given size of 1~$R_{\rm Jup}$, for all effective temperatures between 4,000 and 400 K, are also drawn on Fig. \ref{fig:CM}, with a mark where 800\,K is. WASP-80b coincides with blackbodies {  of temperatures $890^{+80}_{-60}$K and $900^{+60}_{-40}$K for [3.6] and [4.5 $\mu$m] respectively. These temperatures are equivalent to brightness temperatures but are estimated differently than the values provided in Table~\ref{tab:res}. In this new approach, we make no assumption on the shape of the stellar photosphere, but instead are dependent on the assumed distance. It is reassuring to see agreement between these two different methods.}.

In the [3.6]-[4.5] colour, the T-dwarf sequence and the blackbody both slope toward redder colours. This is in contrast to {  the near IR $J$, $H$ \& $K_{\rm s}$ and mid-IR [5.8 $\mu$m]} bands where the L-T transition is characterised by a blueward slope (while the blackbody remains redward; {  \citet{Triaud:2014ly}).  Predictions are presented in Table~\ref{tab:prediction}. Observing in the $J$, $H$ and $K_{\rm s}$ bands is particularly promising to distinguish between both propositions.}

The brown dwarf sequences and the blackbody expectations are furthest apart, over the two {\it Warm Spitzer} channels, for planets with equilibrium temperatures intermediate between WASP-80b and 1,000K (e.g. WASP-8b; \citealt{Cubillos:2013kx}). Observations of additional systems in this range will help interpret the results for WASP-80b.  Those observations would also verify the existence of diversifying atmospheric properties with increasing magnitude, as revealed by an increase in the range of colour occupied by cooler planets \citep{Triaud:2014ly}.

\subsection{About the transmission spectrum}\label{sec:transmi}

\begin{figure*}
\center
\includegraphics[width= 0.8\textwidth]{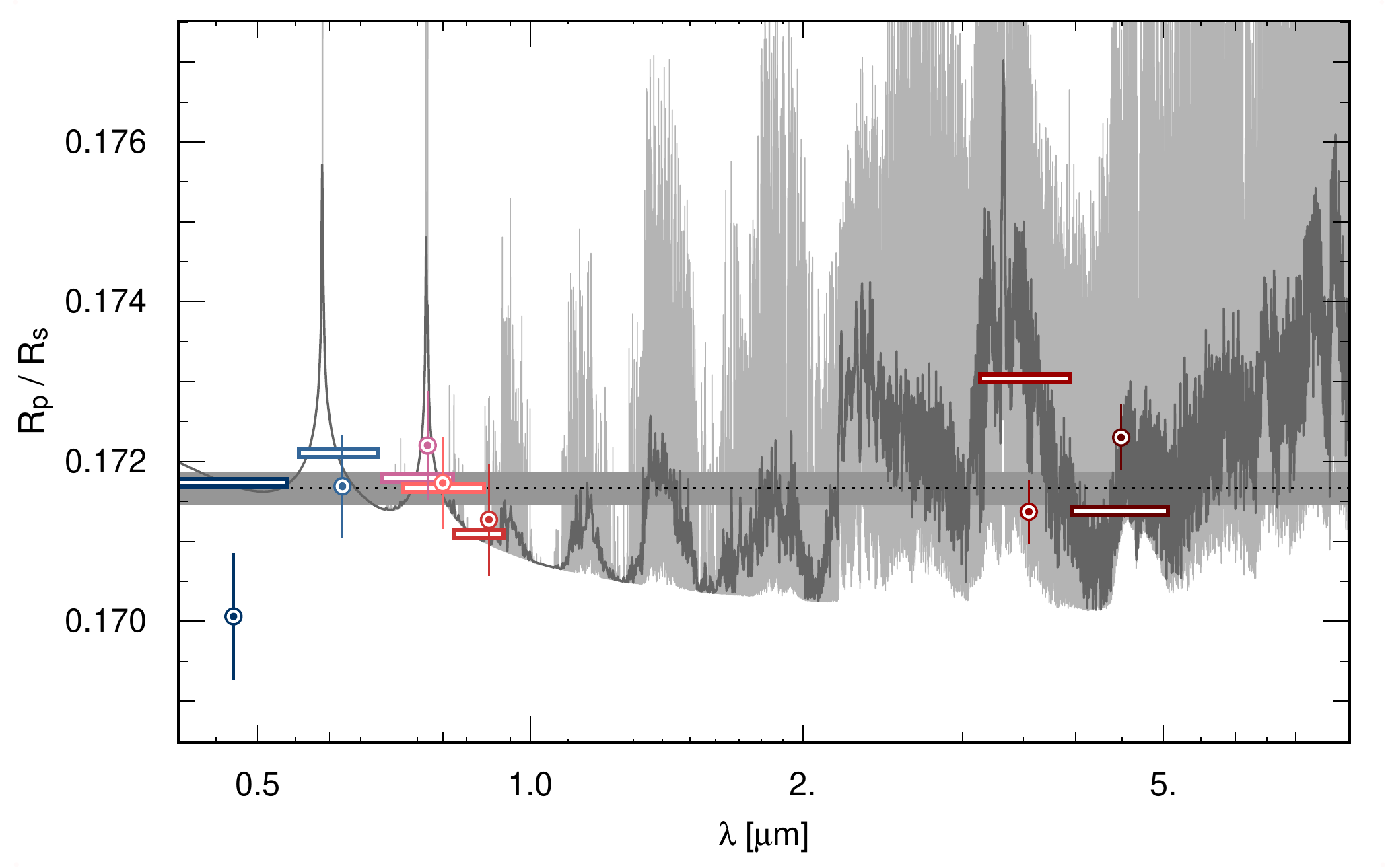}
\caption{Ratios of the planet to star radii as a function of wavelength. The dotted line is the weighed mean, and the grey area the error on that mean. A model atmosphere (H$_2$: 93\% , He: 7\%, H$_2$O: $10^{-4}$ , CH$_4$: $10^{-4}$ , CO: $10^{-4}$ , Na: $10^{-6}$  and K: $10^{-7}$) is also displayed, with two different resolutions, drawn using different shades. This model is not fitted, and is here purely for illustrative purposes. {  Coloured bars correspond to the measurements as expected from the model; their widths illustrate the full width at half maximum of their respective filters.}
}
\label{fig:transmi}
\end{figure*}

The planet-star radius ratio ($R_{\rm p} / R_\star$) in each of our broad bandpasses can be found at the bottom of Table~\ref{tab:res}. They are graphically presented in Fig.~\ref{fig:transmi}, and except for the value at 0.47 $\mu$m, are all consistent with each other. The inverse-variance weighted average of these values is $R_{\rm p} / R_\star = 0.17167\pm 0.00020$, which is also plotted on Fig.~\ref{fig:transmi}. 
The data agree well with a horizontal line: $\chi^2 = 8.3\pm4.1$, which for six degrees of freedom (seven wavelengths and one free parameter) makes a reduced $\chi^2_{\rm r} = 1.4\pm0.7$. We see no reason to compare our data to more complex models (even a slope would be too much). Our analysis refines the results of \citet{Mancini:2014lr}. WASP-80b joins a number of hot Jupiters whose spectra are approximately flat within measurement errors (e.g. \citealt{Pont:2013lr, Mancini:2013fr}). Flat transmission spectra have also been observed for warm Neptunes such as GJ\,1214b \citep{Berta:2012fk,Kreidberg:2014xy}, GJ\,436b \citep{Knutson:2014fk}, GJ\,3470b \citep{Ehrenreich:2014jk}, a result that has been interpreted as indicating the presence of clouds {  or as a sign of high metal content.}

\citet{Fukui:2014bh} also observed WASP-80b. They find $R_{\rm p} / R_\star = 0.17000\pm0.00040$ (averaged over the $J$, $H$ and $K_{\rm s}$ bands, who are all consistent with each other) a value similar to the measurement at 0.45 $\mu m$ (Sloan g') shown in Fig.~\ref{fig:transmi}. In the visible they measure $R_{\rm p} / R_\star = 0.1741\pm0.0010$, which is higher than what we find\footnote{those values only account for their transit depths, after removing those of \citet{Mancini:2014lr} which they had included too.}. In their analysis, some transit parameters were kept fixed which usually results in underestimated error bars and would make any dispersion appear statistically more significant than it really is. For this reason, we have not included their results in our analysis. 

If the redness of WASP-80b's dayside is an indication of methane absorption then a corresponding signature is to be expected in transmission. \cite{Fortney:2010qd} present models for planets with equilibrium temperatures of 1000\,K and  500\,K. While the former remains mostly influenced by H$_2$O, Na and K absorption, the latter shows a transition to an atmosphere dominated by CH$_4$ absorption, closer to what we expect in the case of WASP-80b. {  This simple picture can be considerably altered in the presence of an opaque cloud cover (possibly assembled from soots, created through via photochemical reactions with methane \citep{Moses:2011lr}).}

For illustrative purposes, we calculated a model spectrum for WASP-80b, which we display in Fig. \ref{fig:transmi} \citep{Ehrenreich:2006kx,Ehrenreich:2014jk}\footnote{nota bene: this model does not solve to radiative equilibrium}. It was set with the following abundances: {  H$_2$: 93\% , He: 7\%, H$_2$O: $10^{-4}$, CH$_4$: $10^{-4}$, CO: $10^{-4}$, Na: $10^{-6}$  and K: $10^{-7}$ as in \citet{Moses:2011lr}}. {  The visible regions are dominated by alkali absorption, with some Rayleigh scattering, produced purely from H$_2$}. We note that the bluest measurement is particularly discrepant to that model. A visual comparison of that model to our data also shows that one of the strongest methane features peaks around 3.5 $\mu$m. Methane should enhance a planet's radius at [3.6] compared to [4.5], which we do not see in transmission ({  although} we detect an equivalent effect in emission). This is the most we can reliably affirm. 


\section{Conclusion}

WASP-80b is on an aligned orbit and is the second faintest planet with an emission measured in each of the {\it Warm Spitzer} channels, after GJ\,436b at [3.6] \citep{Lanotte:2014xy} and 55 Cnc e at [4.5] \citep{Demory:2012uq}, which are indicated in Fig.~\ref{fig:CM}. It is the faintest measured in both bands and for which a colour can be computed.  The planet's dayside is amongst the three reddest daysides of any hot Jupiter. HD\,189733Ab's nightside remains the measurement furthest to the red \citep{Knutson:2012ys,Triaud:2014kq}.

GJ\,436b and WASP-80b have similar equilibrium temperatures \citep{Stevenson:2010oq} {  and both show a flat spectrum \citep{Knutson:2014fk}. However, we find that their colours are radically different which can be interpreted as evidence for different chemical compositions (as proposed in \citet{Stevenson:2010oq,Lanotte:2014xy}).} We note here that GJ 436b has a radius significantly smaller than WASP-80b's; one could wonder if both ought to be compared since gravity has an important role to play on the condensation of chemical species into clouds layers (GJ\,436b is 2.3 times denser than WASP-80b). However, we find that there is some credence to that:  a change in radius will cause a change in absolute magnitude. Correcting GJ\,436b's radius to WASP-80b's (1~$R_{\rm Jup}$) amounts to comparing absolute magnitude produced by similar areas. If it were that large, GJ\,436b would be 2.1 magnitude brighter. Using the values presented in \citet{Triaud:2014ly}, this would mean a $M_{[3.6]} = 13.13\pm0.28$, on a par with WASP-80b. Both planets emit a similar flux density at 3.6 $\mu$m (while remaining very different at 4.5 $\mu$m).

{  WASP-80b's atmosphere is ambiguous.} While WASP-80b presents a dayside with a luminosity and a colour reminiscent of a mid T-dwarf (who have a patchy cloud cover and methane absorption), its transmission spectrum appears featureless (implying overcast, without direct evidence for methane). {  The planet is also entirely compatible with a blackbody. Upfront the current data would fit well with WASP-80b being a dusty isothermally radiating ball. However, some brown dwarfs too are compatible with blackbody colours and would too present a flat transmission. Instead, direct spectroscopy shows they are far from resembling blackbodies. We need to keep entertaining the possibility that neither is WASP-80b.}

{  Is WASP-80b a T4? Likely not. If we have learned anything about exoplanets, is that they beat our simplest explanations. However its current position in the colour-magnitude diagram permits us to raise interesting questions. These diagrams become a reading grid.} More observations are dearly needed to first, verify WASP-80b's colour, especially in the 3.6 $\mu$m band where our detection is only at the 4$\sigma$ level and secondly, to scrutinise the planet's size as a function of wavelength in the near infrared, in search for some weak methane absorption.  {  Alternatively, occultation measurements in the near infrared would confirm whether the planet obeys a T dwarf behaviour, constraint its flux to a blackbody, or uncover its own planetary behaviour.}

The methods developed by \citet{Snellen:2010lr} to detect CO and H$_2$O in absorption and in emission in exoplanets (e.g. \citealt{Rodler:2012qy,Birkby:2013vn}) could in principle be used to probe for the presence of methane. Observations have been attempted at the VLT using CRIRES targeting a region containing a forest of CH$_4$ lines, around 2.3 $\mu$m. However bad weather and scheduling issues meant that fewer transits were obtained than requested. CRIRES has now been un-mounted from the telescope for upgrade.

Would a transmission spectrum turn out to show no spectral signatures, we could postulate the following: 
\begin{itemize}
\item WASP-80b is a featureless blackbody, presumably {  because of a dusty/cloudy atmosphere}, that just happened to coincide with the T-dwarf sequence.
\item WASP-80b is part of a diverse population of {\it warm} planets {  enveloped by a greater variety of atmospheres than brown dwarfs have,} as proposed in \citet{Triaud:2014ly}. WASP-80b's position at the conjunction between blackbody and T-dwarfs would therefore be a coincidence. If indeed the case, it implies that its composition cannot be inferred from comparing with ultra-cool dwarfs, such as T-dwarfs, whose {  atmospheric chemistry and optical properties} would share {  little} in common with planets {  (e.g. \citet{Zahnle:2014fk}), especially the irradiated kind}.
\item WASP-80b's colour is an integrated view of its dayside surface which may be in-homogenous (like Kepler-7b; \citealt{Demory:2013uq}). The hottest parts are possibly covered in clouds (like L dwarfs and early type T-dwarfs), while the colder zones of the dayside may remain free, clouds having rained out. Equatorial wind could move enough of the clouds to the night side, traversing the terminator and affecting the resulting transmitted spectrum.
\end{itemize}

To complicate further, the range of colour indices covered by the colder gas giants may be produced by clouds covers having distinct optical properties, for instance, if trace elements specific to some planet react with the surfaces of suspended cloud particles, leading to different albedos and opacities. In addition, extra diversity can be caused by planets possessing a different number and alternation of {\it belts} and {\it zones} with contrasting albedos (Jupiter and Saturn are dissimilar). 

{  Most of these considerations rely on the fact that brown dwarfs and planets can be compared to each others directly. The exercise is currently useful since the brown dwarfs are well understood and that the planets are not. Eventually, we will come to study each population in isolation once planetary-specific behaviour has been identified. Reasons range from the irradiated conditions of hot Jupiters, to being in a different gravity range compared to brown dwarfs. Chemical transitions very likely differ between brown dwarfs and planets. Gravity influences the chemical balance between CO and CH$_4$ \citep{Zahnle:2014fk}. This would blunt the sharp L-T transition and cause a shallower slope in the [3.6]-[4.5] colour than for brown dwarfs, making it closer to a blackbody.}

{  Currently, it remains unclear whether WASP-80b's atmosphere contains any methane. Few planets have been observed near the L/T transition. Apart from WASP-80b, most are young, directly imaged systems which despite distinctive physical parameters, present similar atmospheric characteristics: they are close to blackbodies (in near IR; \citet{Triaud:2014ly}), likely cloudy \citep{Marley:2012fk}, and offer methane absorption lower than expected \citep{Konopacky:2013rt}. As such, deepening the study of WASP-80b will prove important. Parallel properties between the massive, far flung giants orbiting a young A star, and the light, mildly irradiated giant orbiting an M dwarf would inform us about widely occurring phenomena in exoplanets (be they related to atmospheric physics or to formation processes).
}

Doubtless to say, other systems intermediate between the general population of hot Jupiters and WASP-80b would be fascinating to observe, in order to study where they land in colour-magnitude diagrams. Observing the shape of their transmitted spectra can bring to the fore a relation between colour and chemical composition/clouds presence.

\section*{Nota Bene}
Dates are given in the BJD-UTC standard. The radii we used for Jupiter and the Sun are the volumetric mean radii.

\section*{Acknowledgments}

A. H.\,M.\,J. Triaud is a Swiss National Science Foundation (SNSF) fellow under grant number P300P2-147773. M. Gillon and E. Jehin are Research Associates at the F.R.S-FNRS; L. Delrez received the support the support of the F.R.I.A. fund of the FNRS. 
D.E., K. H., and S.U. acknowledge the financial support of the SNSF in the frame of the National Centre for Competence in Research ÔPlanetSÕ.  
E.H. and I.R. acknowledge support from the Spanish Ministry of Economy
and Competitiveness (MINECO) and the ``Fondo Europeo de Desarrollo
Regiona'l' (FEDER) through grants AYA2012-39612-C03-01 and
ESP2013-48391-C4-1-R.

TRAPPIST is a project funded by the Belgian Fund for Scientific Research (Fond National de la Recherche Scientifique, 
F.R.S-FNRS) under grant FRFC 2.5.594.09.F, with the participation of the Swiss National Science Fundation (SNF). 
The {\it Euler} Swiss telescope is supported by the SNSF.
We are all very grateful to ESO and its La Silla  staff for their continuous support. 

This publication makes use of data products from the following projects, whose data was obtained through \href{http://simbad.u-strasbg.fr/simbad/}{Simbad} and \href{http://vizier.u-strasbg.fr/viz-bin/VizieR}{VizieR} services hosted at the \href{http://cds.u-strasbg.fr}{CDS-Strasbourg}:
We gathered the {\it Spitzer} Space Telescope data from the \href{http://sha.ipac.caltech.edu/applications/Spitzer/SHA/}{{\it Spitzer} Heritage Archive}. In addition, this publication has made use of data products from the Wide-field Infrared Survey Explorer ({\it WISE}), which is a joint project of the University of California, Los Angeles, and the Jet Propulsion Laboratory/California Institute of Technology, funded by the National Aeronautics and Space Administration. 
We also collected information from the Two Micron All Sky Survey (2MASS), which is a joint project of the University of Massachusetts and the Infrared Process- ing and Analysis Center/California Institute of Technology, funded by the National Aeronautics and Space Administration and the National Science Foundation.

References to exoplanetary systems were obtained through the use of the paper repositories, \href{http://adsabs.harvard.edu/abstract_service.html}{ADS} and \href{http://arxiv.org/archive/astro-ph}{arXiv}, but also through frequent visits to the \href{http://exoplanet.eu}{exoplanet.eu} \citep{Schneider:2011lr} and \href{http://exoplanets.org}{exoplanets.org} \citep{Wright:2011fj} websites.

\bibliographystyle{mn2e}
\bibliography{1Mybib.bib}

\begin{thebibliography}{84}
\expandafter\ifx\csname natexlab\endcsname\relax\def\natexlab#1{#1}\fi

\bibitem[{{Ackerman} \& {Marley}(2001)}]{Ackerman:2001fk}
{Ackerman} A.~S., {Marley} M.~S., 2001, \apj, 556, 872

\bibitem[{{Albrecht} {et~al}\mbox{.}(2011){Albrecht}, {Winn}, {Johnson},
  {Butler}, {Crane}, {Shectman}, {Thompson}, {Narita}, {Sato}, {Hirano},
  {Enya}, \& {Fischer}}]{Albrecht:2011rt}
{Albrecht} S. {et~al.}, 2011, \apj, 738, 50

\bibitem[{{Artigau} {et~al}\mbox{.}(2009){Artigau}, {Bouchard}, {Doyon}, \&
  {Lafreni{\`e}re}}]{Artigau:2009lr}
{Artigau} {\'E}., {Bouchard} S., {Doyon} R., {Lafreni{\`e}re} D., 2009, \apj,
  701, 1534

\bibitem[{{Baraffe} {et~al}\mbox{.}(2003){Baraffe}, {Chabrier}, {Barman},
  {Allard}, \& {Hauschildt}}]{Baraffe:2003gf}
{Baraffe} I., {Chabrier} G., {Barman} T.~S., {Allard} F., {Hauschildt} P.~H.,
  2003, \aap, 402, 701

\bibitem[{{Berta} {et~al}\mbox{.}(2012){Berta}, {Charbonneau}, {D{\'e}sert},
  {Miller-Ricci Kempton}, {McCullough}, {Burke}, {Fortney}, {Irwin}, {Nutzman},
  \& {Homeier}}]{Berta:2012fk}
{Berta} Z.~K. {et~al.}, 2012, \apj, 747, 35

\bibitem[{{Birkby} {et~al}\mbox{.}(2013){Birkby}, {de Kok}, {Brogi}, {de
  Mooij}, {Schwarz}, {Albrecht}, \& {Snellen}}]{Birkby:2013vn}
{Birkby} J.~L., {de Kok} R.~J., {Brogi} M., {de Mooij} E.~J.~W., {Schwarz} H.,
  {Albrecht} S., {Snellen} I.~A.~G., 2013, \mnras, 436, L35

\bibitem[{{Bolton} \& {the Juno Science Team}(2010)}]{Bolton:2010xy}
{Bolton} S.~J., {the Juno Science Team}, 2010, in IAU Symposium, Vol. 269, IAU
  Symposium, {Barbieri} C., {Chakrabarti} S., {Coradini} M., {Lazzarin} M.,
  eds., pp. 92--100

\bibitem[{{Buenzli} {et~al}\mbox{.}(2014){Buenzli}, {Apai}, {Radigan}, {Reid},
  \& {Flateau}}]{Buenzli:2014fj}
{Buenzli} E., {Apai} D., {Radigan} J., {Reid} I.~N., {Flateau} D., 2014, \apj,
  782, 77

\bibitem[{{Burgasser} {et~al}\mbox{.}(2002){Burgasser}, {Marley}, {Ackerman},
  {Saumon}, {Lodders}, {Dahn}, {Harris}, \& {Kirkpatrick}}]{Burgasser:2002kx}
{Burgasser} A.~J., {Marley} M.~S., {Ackerman} A.~S., {Saumon} D., {Lodders} K.,
  {Dahn} C.~C., {Harris} H.~C., {Kirkpatrick} J.~D., 2002, \apjl, 571, L151

\bibitem[{{Burrows}(2014)}]{Burrows:2014yu}
{Burrows} A.~S., 2014, Proceedings of the National Academy of Science, 111,
  12601

\bibitem[{{Claret} \& {Bloemen}(2011)}]{Claret:2011rm}
{Claret} A., {Bloemen} S., 2011, \aap, 529, A75

\bibitem[{{Collier Cameron} {et~al}\mbox{.}(2007){Collier Cameron}, {Bouchy},
  {H{\'e}brard}, {Maxted}, {Pollacco}, {Pont}, {Skillen}, {Smalley}, {Street},
  {West}, {Wilson}, {Aigrain}, {Christian}, {Clarkson}, {Enoch}, {Evans},
  {Fitzsimmons}, {Fleenor}, {Gillon}, {Haswell}, {Hebb}, {Hellier}, {Hodgkin},
  {Horne}, {Irwin}, {Kane}, {Keenan}, {Loeillet}, {Lister}, {Mayor}, {Moutou},
  {Norton}, {Osborne}, {Parley}, {Queloz}, {Ryans}, {Triaud}, {Udry}, \&
  {Wheatley}}]{Collier-Cameron:2007fj}
{Collier Cameron} A. {et~al.}, 2007, \mnras, 375, 951

\bibitem[{{Cooper} \& {Showman}(2006)}]{Cooper:2006yu}
{Cooper} C.~S., {Showman} A.~P., 2006, \apj, 649, 1048

\bibitem[{{Crossfield} {et~al}\mbox{.}(2014){Crossfield}, {Biller},
  {Schlieder}, {Deacon}, {Bonnefoy}, {Homeier}, {Allard}, {Buenzli}, {Henning},
  {Brandner}, {Goldman}, \& {Kopytova}}]{Crossfield:2014db}
{Crossfield} I.~J.~M. {et~al.}, 2014, \nat, 505, 654

\bibitem[{{Cubillos} {et~al}\mbox{.}(2013){Cubillos}, {Harrington},
  {Madhusudhan}, {Stevenson}, {Hardy}, {Blecic}, {Anderson}, {Hardin}, \&
  {Campo}}]{Cubillos:2013kx}
{Cubillos} P. {et~al.}, 2013, \apj, 768, 42

\bibitem[{{Cutri} \& {et al.}(2013)}]{Cutri:2013fj}
{Cutri} R.~M., {et al.}, 2013, VizieR Online Data Catalog, 2328, 0

\bibitem[{{Cutri} {et~al}\mbox{.}(2003){Cutri}, {Skrutskie}, {van Dyk},
  {Beichman}, {Carpenter}, {Chester}, {Cambresy}, {Evans}, {Fowler}, {Gizis},
  {Howard}, {Huchra}, {Jarrett}, {Kopan}, {Kirkpatrick}, {Light}, {Marsh},
  {McCallon}, {Schneider}, {Stiening}, {Sykes}, {Weinberg}, {Wheaton},
  {Wheelock}, \& {Zacarias}}]{Cutri:2003rt}
{Cutri} R.~M. {et~al.}, 2003, VizieR Online Data Catalog, 2246, 0

\bibitem[{{Demory} {et~al}\mbox{.}(2013){Demory}, {de Wit}, {Lewis}, {Fortney},
  {Zsom}, {Seager}, {Knutson}, {Heng}, {Madhusudhan}, {Gillon}, {Barclay},
  {Desert}, {Parmentier}, \& {Cowan}}]{Demory:2013uq}
{Demory} B.-O. {et~al.}, 2013, \apjl, 776, L25

\bibitem[{{Demory} {et~al}\mbox{.}(2012){Demory}, {Gillon}, {Seager},
  {Benneke}, {Deming}, \& {Jackson}}]{Demory:2012uq}
{Demory} B.-O., {Gillon} M., {Seager} S., {Benneke} B., {Deming} D., {Jackson}
  B., 2012, \apjl, 751, L28

\bibitem[{{Demory} \& {Seager}(2011)}]{Demory:2011lr}
{Demory} B.-O., {Seager} S., 2011, \apjs, 197, 12

\bibitem[{{Doyle} {et~al}\mbox{.}(2014){Doyle}, {Davies}, {Smalley}, {Chaplin},
  \& {Elsworth}}]{Doyle:2014qf}
{Doyle} A.~P., {Davies} G.~R., {Smalley} B., {Chaplin} W.~J., {Elsworth} Y.,
  2014, \mnras, 444, 3592

\bibitem[{{Dupuy} \& {Liu}(2012)}]{Dupuy:2012lr}
{Dupuy} T.~J., {Liu} M.~C., 2012, \apjs, 201, 19

\bibitem[{{Ehrenreich} {et~al}\mbox{.}(2014){Ehrenreich}, {Bonfils}, {Lovis},
  {Delfosse}, {Forveille}, {Mayor}, {Neves}, {Santos}, {Udry}, \&
  {S{\'e}gransan}}]{Ehrenreich:2014jk}
{Ehrenreich} D. {et~al.}, 2014, \aap, 570, A89

\bibitem[{{Ehrenreich} {et~al}\mbox{.}(2006){Ehrenreich}, {Tinetti},
  {Lecavelier Des Etangs}, {Vidal-Madjar}, \& {Selsis}}]{Ehrenreich:2006kx}
{Ehrenreich} D., {Tinetti} G., {Lecavelier Des Etangs} A., {Vidal-Madjar} A.,
  {Selsis} F., 2006, \aap, 448, 379

\bibitem[{{Fazio} {et~al}\mbox{.}(2004){Fazio}, {Hora}, {Allen}, {Ashby},
  {Barmby}, {Deutsch}, {Huang}, {Kleiner}, {Marengo}, {Megeath}, {Melnick},
  {Pahre}, {Patten}, {Polizotti}, {Smith}, {Taylor}, {Wang}, {Willner},
  {Hoffmann}, {Pipher}, {Forrest}, {McMurty}, {McCreight}, {McKelvey},
  {McMurray}, {Koch}, {Moseley}, {Arendt}, {Mentzell}, {Marx}, {Losch},
  {Mayman}, {Eichhorn}, {Krebs}, {Jhabvala}, {Gezari}, {Fixsen}, {Flores},
  {Shakoorzadeh}, {Jungo}, {Hakun}, {Workman}, {Karpati}, {Kichak}, {Whitley},
  {Mann}, {Tollestrup}, {Eisenhardt}, {Stern}, {Gorjian}, {Bhattacharya},
  {Carey}, {Nelson}, {Glaccum}, {Lacy}, {Lowrance}, {Laine}, {Reach},
  {Stauffer}, {Surace}, {Wilson}, {Wright}, {Hoffman}, {Domingo}, \&
  {Cohen}}]{Fazio:2004fk}
{Fazio} G.~G. {et~al.}, 2004, \apjs, 154, 10

\bibitem[{{Fortney} {et~al}\mbox{.}(2010){Fortney}, {Shabram}, {Showman},
  {Lian}, {Freedman}, {Marley}, \& {Lewis}}]{Fortney:2010qd}
{Fortney} J.~J., {Shabram} M., {Showman} A.~P., {Lian} Y., {Freedman} R.~S.,
  {Marley} M.~S., {Lewis} N.~K., 2010, \apj, 709, 1396

\bibitem[{{Fraine} {et~al}\mbox{.}(2014){Fraine}, {Deming}, {Benneke},
  {Knutson}, {Jord{\'a}n}, {Espinoza}, {Madhusudhan}, {Wilkins}, \&
  {Todorov}}]{Fraine:2014lr}
{Fraine} J. {et~al.}, 2014, \nat, 513, 526

\bibitem[{{Fukui} {et~al}\mbox{.}(2014){Fukui}, {Kawashima}, {Ikoma}, {Narita},
  {Onitsuka}, {Ita}, {Onozato}, {Nishiyama}, {Baba}, {Ryu}, {Hirano}, {Hori},
  {Kurosaki}, {Kawauchi}, {Takahashi}, {Nagayama}, {Tamura}, {Kawai}, {Kuroda},
  {Nagayama}, {Ohta}, {Shimizu}, {Yanagisawa}, {Yoshida}, \&
  {Izumiura}}]{Fukui:2014bh}
{Fukui} A. {et~al.}, 2014, \apj, 790, 108

\bibitem[{{Gelman} \& {Rubin}(1992)}]{Gelman:1992rt}
{Gelman} A., {Rubin} D.~B., 1992, Statitical Science, 7, 457

\bibitem[{{Gillon} {et~al}\mbox{.}(2014){Gillon}, {Demory}, {Madhusudhan},
  {Deming}, {Seager}, {Zsom}, {Knutson}, {Lanotte}, {Bonfils}, {D{\'e}sert},
  {Delrez}, {Jehin}, {Fraine}, {Magain}, \& {Triaud}}]{Gillon:2014ix}
{Gillon} M. {et~al.}, 2014, \aap, 563, A21

\bibitem[{{Gillon} {et~al}\mbox{.}(2012){Gillon}, {Triaud}, {Fortney},
  {Demory}, {Jehin}, {Lendl}, {Magain}, {Kabath}, {Queloz}, {Alonso},
  {Anderson}, {Collier Cameron}, {Fumel}, {Hebb}, {Hellier}, {Lanotte},
  {Maxted}, {Mowlavi}, \& {Smalley}}]{Gillon:2012fj}
---, 2012, \aap, 542, A4

\bibitem[{{Gillon} {et~al}\mbox{.}(2013){Gillon}, {Triaud}, {Jehin}, {Delrez},
  {Opitom}, {Magain}, {Lendl}, \& {Queloz}}]{Gillon:2013qv}
{Gillon} M., {Triaud} A.~H.~M.~J., {Jehin} E., {Delrez} L., {Opitom} C.,
  {Magain} P., {Lendl} M., {Queloz} D., 2013, \aap, 555, L5

\bibitem[{{Gim{\'e}nez}(2006)}]{Gimenez:2006kx}
{Gim{\'e}nez} A., 2006, \apj, 650, 408

\bibitem[{{Gray}(2008)}]{Gray:2008fj}
{Gray} D.~F., 2008, {The Observation and Analysis of Stellar Photospheres},
  {Gray, D.~F.}, ed.

\bibitem[{{Hellier} {et~al}\mbox{.}(2012){Hellier}, {Anderson}, {Collier
  Cameron}, {Doyle}, {Fumel}, {Gillon}, {Jehin}, {Lendl}, {Maxted}, {Pepe},
  {Pollacco}, {Queloz}, {S{\'e}gransan}, {Smalley}, {Smith}, {Southworth},
  {Triaud}, {Udry}, \& {West}}]{Hellier:2012fj}
{Hellier} C. {et~al.}, 2012, \mnras, 426, 739

\bibitem[{{Herrero}, {Ribas} \& {Jordi}(2014){Herrero}, {Ribas}, \&
  {Jordi}}]{Herrero:2014lr}
{Herrero} E., {Ribas} I., {Jordi} C., 2014, Experimental Astronomy

\bibitem[{{Jehin} {et~al}\mbox{.}(2011){Jehin}, {Gillon}, {Queloz}, {Magain},
  {Manfroid}, {Chantry}, {Lendl}, {Hutsem{\'e}kers}, \& {Udry}}]{Jehin:2011dk}
{Jehin} E. {et~al.}, 2011, The Messenger, 145, 2

\bibitem[{{Jord{\'a}n} {et~al}\mbox{.}(2013){Jord{\'a}n}, {Espinoza}, {Rabus},
  {Eyheramendy}, {Sing}, {D{\'e}sert}, {Bakos}, {Fortney}, {L{\'o}pez-Morales},
  {Maxted}, {Triaud}, \& {Szentgyorgyi}}]{Jordan:2013dn}
{Jord{\'a}n} A. {et~al.}, 2013, \apj, 778, 184

\bibitem[{{Kirkpatrick}(2005)}]{Kirkpatrick:2005th}
{Kirkpatrick} J.~D., 2005, \araa, 43, 195

\bibitem[{{Knutson} {et~al}\mbox{.}(2014){Knutson}, {Benneke}, {Deming}, \&
  {Homeier}}]{Knutson:2014fk}
{Knutson} H.~A., {Benneke} B., {Deming} D., {Homeier} D., 2014, \nat, 505, 66

\bibitem[{{Knutson} {et~al}\mbox{.}(2008){Knutson}, {Charbonneau}, {Allen},
  {Burrows}, \& {Megeath}}]{Knutson:2008qy}
{Knutson} H.~A., {Charbonneau} D., {Allen} L.~E., {Burrows} A., {Megeath}
  S.~T., 2008, \apj, 673, 526

\bibitem[{{Knutson} {et~al}\mbox{.}(2012){Knutson}, {Lewis}, {Fortney},
  {Burrows}, {Showman}, {Cowan}, {Agol}, {Aigrain}, {Charbonneau}, {Deming},
  {D{\'e}sert}, {Henry}, {Langton}, \& {Laughlin}}]{Knutson:2012ys}
{Knutson} H.~A. {et~al.}, 2012, \apj, 754, 22

\bibitem[{{Konopacky} {et~al}\mbox{.}(2013){Konopacky}, {Barman}, {Macintosh},
  \& {Marois}}]{Konopacky:2013rt}
{Konopacky} Q.~M., {Barman} T.~S., {Macintosh} B.~A., {Marois} C., 2013,
  Science, 339, 1398

\bibitem[{{Kreidberg} {et~al}\mbox{.}(2014){Kreidberg}, {Bean}, {D{\'e}sert},
  {Benneke}, {Deming}, {Stevenson}, {Seager}, {Berta-Thompson}, {Seifahrt}, \&
  {Homeier}}]{Kreidberg:2014xy}
{Kreidberg} L. {et~al.}, 2014, \nat, 505, 69

\bibitem[{{Kurucz}(1993)}]{Kurucz:1993qy}
{Kurucz} R., 1993, ATLAS9 Stellar Atmosphere Programs and 2 km/s grid.~Kurucz
  CD-ROM No.~13.~ Cambridge, Mass.: Smithsonian Astrophysical Observatory,
  1993., 13

\bibitem[{{Lanotte} {et~al}\mbox{.}(2014){Lanotte}, {Gillon}, {Demory},
  {Fortney}, {Astudillo}, {Bonfils}, {Magain}, {Delfosse}, {Forveille},
  {Lovis}, {Mayor}, {Neves}, {Pepe}, {Queloz}, {Santos}, \&
  {Udry}}]{Lanotte:2014xy}
{Lanotte} A.~A. {et~al.}, 2014, \aap, 572, A73

\bibitem[{{Lewis} {et~al}\mbox{.}(2013){Lewis}, {Knutson}, {Showman}, {Cowan},
  {Laughlin}, {Burrows}, {Deming}, {Crepp}, {Mighell}, {Agol}, {Bakos},
  {Charbonneau}, {D{\'e}sert}, {Fischer}, {Fortney}, {Hartman}, {Hinkley},
  {Howard}, {Johnson}, {Kao}, {Langton}, \& {Marcy}}]{Lewis:2013qy}
{Lewis} N.~K. {et~al.}, 2013, \apj, 766, 95

\bibitem[{{Lin}, {Bodenheimer} \& {Richardson}(1996){Lin}, {Bodenheimer}, \&
  {Richardson}}]{Lin:1996yq}
{Lin} D.~N.~C., {Bodenheimer} P., {Richardson} D.~C., 1996, \nat, 380, 606

\bibitem[{{L{\'o}pez-Morales} {et~al}\mbox{.}(2014){L{\'o}pez-Morales},
  {Triaud}, {Rodler}, {Dumusque}, {Buchhave}, {Harutyunyan}, {Hoyer}, {Alonso},
  {Gillon}, {Kaib}, {Latham}, {Lovis}, {Pepe}, {Queloz}, {Raymond},
  {S{\'e}gransan}, {Waldmann}, \& {Udry}}]{Lopez-Morales:2014qv}
{L{\'o}pez-Morales} M. {et~al.}, 2014, \apjl, 792, L31

\bibitem[{{Madhusudhan} {et~al}\mbox{.}(2014){Madhusudhan}, {Crouzet},
  {McCullough}, {Deming}, \& {Hedges}}]{Madhusudhan:2014jk}
{Madhusudhan} N., {Crouzet} N., {McCullough} P.~R., {Deming} D., {Hedges} C.,
  2014, \apjl, 791, L9

\bibitem[{{Mancini} {et~al}\mbox{.}(2013){Mancini}, {Nikolov}, {Southworth},
  {Chen}, {Fortney}, {Tregloan-Reed}, {Ciceri}, {van Boekel}, \&
  {Henning}}]{Mancini:2013fr}
{Mancini} L. {et~al.}, 2013, \mnras, 430, 2932

\bibitem[{{Mancini} {et~al}\mbox{.}(2014){Mancini}, {Southworth}, {Ciceri},
  {Dominik}, {Henning}, {J{\o}rgensen}, {Lanza}, {Rabus}, {Snodgrass},
  {Vilela}, {Alsubai}, {Bozza}, {Bramich}, {Calchi Novati}, {D'Ago}, {Figuera
  Jaimes}, {Galianni}, {Gu}, {Harps{\o}e}, {Hinse}, {Hundertmark}, {Juncher},
  {Kains}, {Korhonen}, {Popovas}, {Rahvar}, {Skottfelt}, {Street}, {Surdej},
  {Tsapras}, {Wang}, \& {Wertz}}]{Mancini:2014lr}
---, 2014, \aap, 562, A126

\bibitem[{{Mandel} \& {Agol}(2002)}]{Mandel:2002kx}
{Mandel} K., {Agol} E., 2002, \apjl, 580, L171

\bibitem[{{Mandell} {et~al}\mbox{.}(2013){Mandell}, {Haynes}, {Sinukoff},
  {Madhusudhan}, {Burrows}, \& {Deming}}]{Mandell:2013fk}
{Mandell} A.~M., {Haynes} K., {Sinukoff} E., {Madhusudhan} N., {Burrows} A.,
  {Deming} D., 2013, \apj, 779, 128

\bibitem[{{Marley} {et~al}\mbox{.}(2012){Marley}, {Saumon}, {Cushing},
  {Ackerman}, {Fortney}, \& {Freedman}}]{Marley:2012fk}
{Marley} M.~S., {Saumon} D., {Cushing} M., {Ackerman} A.~S., {Fortney} J.~J.,
  {Freedman} R., 2012, \apj, 754, 135

\bibitem[{{Mayor} \& {Queloz}(1995)}]{Mayor:1995uq}
{Mayor} M., {Queloz} D., 1995, \nat, 378, 355

\bibitem[{{McCullough} {et~al}\mbox{.}(2014){McCullough}, {Crouzet}, {Deming},
  \& {Madhusudhan}}]{McCullough:2014lr}
{McCullough} P.~R., {Crouzet} N., {Deming} D., {Madhusudhan} N., 2014, \apj,
  791, 55

\bibitem[{{Moses} {et~al}\mbox{.}(2011){Moses}, {Visscher}, {Fortney},
  {Showman}, {Lewis}, {Griffith}, {Klippenstein}, {Shabram}, {Friedson},
  {Marley}, \& {Freedman}}]{Moses:2011lr}
{Moses} J.~I. {et~al.}, 2011, \apj, 737, 15

\bibitem[{{{\"O}berg}, {Murray-Clay} \& {Bergin}(2011){{\"O}berg},
  {Murray-Clay}, \& {Bergin}}]{Oberg:2011qv}
{{\"O}berg} K.~I., {Murray-Clay} R., {Bergin} E.~A., 2011, \apjl, 743, L16

\bibitem[{{Patten} {et~al}\mbox{.}(2006){Patten}, {Stauffer}, {Burrows},
  {Marengo}, {Hora}, {Luhman}, {Sonnett}, {Henry}, {Raghavan}, {Megeath},
  {Liebert}, \& {Fazio}}]{Patten:2006uq}
{Patten} B.~M. {et~al.}, 2006, \apj, 651, 502

\bibitem[{{Pollacco} {et~al}\mbox{.}(2006){Pollacco}, {Skillen}, {Collier
  Cameron}, {Christian}, {Hellier}, {Irwin}, {Lister}, {Street}, {West},
  {Anderson}, {Clarkson}, {Deeg}, {Enoch}, {Evans}, {Fitzsimmons}, {Haswell},
  {Hodgkin}, {Horne}, {Kane}, {Keenan}, {Maxted}, {Norton}, {Osborne},
  {Parley}, {Ryans}, {Smalley}, {Wheatley}, \& {Wilson}}]{Pollacco:2006fj}
{Pollacco} D.~L. {et~al.}, 2006, \pasp, 118, 1407

\bibitem[{{Pont} {et~al}\mbox{.}(2013){Pont}, {Sing}, {Gibson}, {Aigrain},
  {Henry}, \& {Husnoo}}]{Pont:2013lr}
{Pont} F., {Sing} D.~K., {Gibson} N.~P., {Aigrain} S., {Henry} G., {Husnoo} N.,
  2013, \mnras, 432, 2917

\bibitem[{{Radigan} {et~al}\mbox{.}(2012){Radigan}, {Jayawardhana},
  {Lafreni{\`e}re}, {Artigau}, {Marley}, \& {Saumon}}]{Radigan:2012rt}
{Radigan} J., {Jayawardhana} R., {Lafreni{\`e}re} D., {Artigau} {\'E}.,
  {Marley} M., {Saumon} D., 2012, \apj, 750, 105

\bibitem[{{Rasio} \& {Ford}(1996)}]{Rasio:1996ly}
{Rasio} F.~A., {Ford} E.~B., 1996, Science, 274, 954

\bibitem[{{Rodler}, {Lopez-Morales} \& {Ribas}(2012){Rodler}, {Lopez-Morales},
  \& {Ribas}}]{Rodler:2012qy}
{Rodler} F., {Lopez-Morales} M., {Ribas} I., 2012, \apjl, 753, L25

\bibitem[{{Schlaufman}(2010)}]{Schlaufman:2010fk}
{Schlaufman} K.~C., 2010, \apj, 719, 602

\bibitem[{{Schneider} {et~al}\mbox{.}(2011){Schneider}, {Dedieu}, {Le Sidaner},
  {Savalle}, \& {Zolotukhin}}]{Schneider:2011lr}
{Schneider} J., {Dedieu} C., {Le Sidaner} P., {Savalle} R., {Zolotukhin} I.,
  2011, \aap, 532, A79

\bibitem[{{Schwarz}(1978)}]{Schwarz:1978zz}
{Schwarz} G., 1978, Annals of Statistics, 6, 461

\bibitem[{{Seager} \& {Deming}(2010)}]{Seager:2010kx}
{Seager} S., {Deming} D., 2010, \araa, 48, 631

\bibitem[{{Sing} {et~al}\mbox{.}(2011){Sing}, {Pont}, {Aigrain}, {Charbonneau},
  {D{\'e}sert}, {Gibson}, {Gilliland}, {Hayek}, {Henry}, {Knutson}, {Lecavelier
  Des Etangs}, {Mazeh}, \& {Shporer}}]{Sing:2011ly}
{Sing} D.~K. {et~al.}, 2011, \mnras, 416, 1443

\bibitem[{{Snellen} {et~al}\mbox{.}(2010){Snellen}, {de Kok}, {de Mooij}, \&
  {Albrecht}}]{Snellen:2010lr}
{Snellen} I.~A.~G., {de Kok} R.~J., {de Mooij} E.~J.~W., {Albrecht} S., 2010,
  \nat, 465, 1049

\bibitem[{{Stephens} {et~al}\mbox{.}(2009){Stephens}, {Leggett}, {Cushing},
  {Marley}, {Saumon}, {Geballe}, {Golimowski}, {Fan}, \&
  {Noll}}]{Stephens:2009lr}
{Stephens} D.~C. {et~al.}, 2009, \apj, 702, 154

\bibitem[{{Stetson}(1987)}]{Stetson:1987kl}
{Stetson} P.~B., 1987, \pasp, 99, 191

\bibitem[{{Stevenson} {et~al}\mbox{.}(2012){Stevenson}, {Harrington},
  {Fortney}, {Loredo}, {Hardy}, {Nymeyer}, {Bowman}, {Cubillos}, {Bowman}, \&
  {Hardin}}]{Stevenson:2012vn}
{Stevenson} K.~B. {et~al.}, 2012, \apj, 754, 136

\bibitem[{{Stevenson} {et~al}\mbox{.}(2010){Stevenson}, {Harrington},
  {Nymeyer}, {Madhusudhan}, {Seager}, {Bowman}, {Hardy}, {Deming}, {Rauscher},
  \& {Lust}}]{Stevenson:2010oq}
---, 2010, \nat, 464, 1161

\bibitem[{{Triaud}(2014)}]{Triaud:2014kq}
{Triaud} A.~H.~M.~J., 2014, \mnras, 439, L61

\bibitem[{{Triaud} {et~al}\mbox{.}(2013){Triaud}, {Anderson}, {Collier
  Cameron}, {Doyle}, {Fumel}, {Gillon}, {Hellier}, {Jehin}, {Lendl}, {Lovis},
  {Maxted}, {Pepe}, {Pollacco}, {Queloz}, {S{\'e}gransan}, {Smalley}, {Smith},
  {Udry}, {West}, \& {Wheatley}}]{Triaud:2013sk}
{Triaud} A.~H.~M.~J. {et~al.}, 2013, \aap, 551, A80

\bibitem[{{Triaud} {et~al}\mbox{.}(2014){Triaud}, {Lanotte}, {Smalley}, \&
  {Gillon}}]{Triaud:2014ly}
{Triaud} A.~H.~M.~J., {Lanotte} A.~A., {Smalley} B., {Gillon} M., 2014, \mnras,
  444, 711

\bibitem[{{Triaud} {et~al}\mbox{.}(2011){Triaud}, {Queloz}, {Hellier},
  {Gillon}, {Smalley}, {Hebb}, {Collier Cameron}, {Anderson}, {Boisse},
  {H{\'e}brard}, {Jehin}, {Lister}, {Lovis}, {Maxted}, {Pepe}, {Pollacco},
  {S{\'e}gransan}, {Simpson}, {Udry}, \& {West}}]{Triaud:2011vn}
{Triaud} A.~H.~M.~J. {et~al.}, 2011, \aap, 531, A24

\bibitem[{{Winn} {et~al}\mbox{.}(2010){Winn}, {Fabrycky}, {Albrecht}, \&
  {Johnson}}]{Winn:2010rr}
{Winn} J.~N., {Fabrycky} D., {Albrecht} S., {Johnson} J.~A., 2010, \apjl, 718,
  L145

\bibitem[{{Wright} {et~al}\mbox{.}(2010){Wright}, {Eisenhardt}, {Mainzer},
  {Ressler}, {Cutri}, {Jarrett}, {Kirkpatrick}, {Padgett}, {McMillan},
  {Skrutskie}, {Stanford}, {Cohen}, {Walker}, {Mather}, {Leisawitz}, {Gautier},
  {McLean}, {Benford}, {Lonsdale}, {Blain}, {Mendez}, {Irace}, {Duval}, {Liu},
  {Royer}, {Heinrichsen}, {Howard}, {Shannon}, {Kendall}, {Walsh}, {Larsen},
  {Cardon}, {Schick}, {Schwalm}, {Abid}, {Fabinsky}, {Naes}, \&
  {Tsai}}]{Wright:2010qv}
{Wright} E.~L. {et~al.}, 2010, \aj, 140, 1868

\bibitem[{{Wright} {et~al}\mbox{.}(2011){Wright}, {Fakhouri}, {Marcy}, {Han},
  {Feng}, {Johnson}, {Howard}, {Fischer}, {Valenti}, {Anderson}, \&
  {Piskunov}}]{Wright:2011fj}
{Wright} J.~T. {et~al.}, 2011, \pasp, 123, 412

\bibitem[{{Zahnle} \& {Marley}(2014)}]{Zahnle:2014fk}
{Zahnle} K.~J., {Marley} M.~S., 2014, \apj, 797, 41

\bibitem[{{Zellem} {et~al}\mbox{.}(2014){Zellem}, {Lewis}, {Knutson},
  {Griffith}, {Showman}, {Fortney}, {Cowan}, {Agol}, {Burrows}, {Charbonneau},
  {Deming}, {Laughlin}, \& {Langton}}]{Zellem:2014rr}
{Zellem} R.~T. {et~al.}, 2014, \apj, 790, 53

\end{thebibliography}

\appendix

\section{}
Here are located the graphical representation of each  photometric time series that has been used in our analysis, including the corrections given to the data. Table~\ref{tab:phot} contains the journal of observations and type of corrections applied to the data.

\begin{table*}
\begin{center}
\scriptsize{
\begin{tabular}{cccccccc}
\hline \noalign {\smallskip} 
Date & Instrument & Filter & $T_{exp}$ &$N_p$ & Baseline & CF               & Eclipse\\
         &                  &           &                   &            & function  &                    &  type      \\     
\hline \noalign {\smallskip} 
7 May 2012  & TRAPPIST & Sloan $z$'  & 10s      & 728   & $p(t^2)+o$                     & 2.08  & transit          \\ \noalign {\smallskip} 
26 Jul 2012    & {\it Euler}    & Gunn $r$'   & 60s    & 208   & $p(t^2+fwhm^2)$           & 1.36  & transit           \\ \noalign {\smallskip} 
10 sep 2012  & TRAPPIST & Sloan $z$'  & 10s     & 705  & $p(t^2)$                          & 1.15  & transit          \\ \noalign {\smallskip} 
13 Jun 2013  & TRAPPIST & Sloan $z$'  & 13s     & 749  & $p(t^2)+o$                      & 1.20  & transit          \\ \noalign {\smallskip} 
16 Jun 2013  & TRAPPIST & Sloan $z$'  & 13s     & 782  & $p(t^2+xy^2)+o$            & 1.49  & transit          \\ \noalign {\smallskip} 
16 Jun 2013   & {\it Euler}    & Gunn $r$'   & 50s    & 175   & $p(t^2)$                            & 2.86  & transit           \\ \noalign {\smallskip} 
16 Jun 2013   & 2.2m/GROND  & Sloan $z$'   & 60s    & 157   & $p(t^2)$                      & 2.64  & transit           \\ \noalign {\smallskip} 
16 Jun 2013   & {\it Danish}/DFOSC  & Bessell $I$  & 60s    & 200   & $p(t^2)$                  & 1.05  & transit           \\ \noalign {\smallskip} 
16 Jun 2013   & 2.2m/GROND  & Sloan $r$'   & 60s    & 156   & $p(t^2)$                       & 2.14  & transit           \\ \noalign {\smallskip} 
16 Jun 2013   & 2.2m/GROND  & Sloan $i$'   & 60s    & 156   & $p(t^2)$                       & 1.05  & transit           \\ \noalign {\smallskip} 
16 Jun 2013   & 2.2m/GROND  & Sloan $g$'   & 60s    & 162  & $p(t^2)$                       & 0.76  & transit           \\ \noalign {\smallskip} 
3 Jul 2013   & {\it Spitzer}  & 3.6 $\mu$m  & & 548   & $p(xy^2+l^1)$ + BM                    & 1.35 & occultation  \\ \noalign {\smallskip} 
13 Jul 2013  & {\it Spitzer}  & 4.5 $\mu$m & & 123   & $p(fwhm_x^1+xy^2+l^1)$           & 1.06 & transit        \\ \noalign {\smallskip} 
18 Jul 2013  & {\it Spitzer} & 4.5 $\mu$m & & 121   & $p(xy^2+l^1)$                              & 0.68 & occultation   \\ \noalign {\smallskip} 
24 Jul 2013 & {\it Spitzer}  & 3.6 $\mu$m & & 569   & $p(fwhm_x^1+xy^2+l^1)$ + BM  & 1.78 & occultation   \\ \noalign {\smallskip} 
27 Jul 2013  & {\it Spitzer}  & 4.5 $\mu$m & & 123   & $p(fwhm_x^1+xy^2+l^1)$           & 1.14 & occultation   \\ \noalign {\smallskip} 
1 Aug 2013  & {\it Spitzer}  & 3.6 $\mu$m & & 579   & $p(fwhm_x^1+xy^2+l^1)$ + BM  & 1.33  & transit          \\ \noalign {\smallskip} 
1 Aug 2013    & TRAPPIST & Sloan $z$'  & 13s    &  512  & $p(t^2)$                             & 1.23  & transit          \\ \noalign {\smallskip} 
\hline
\end{tabular}}
\caption{WASP-80\,b photometric eclipse time-series used in this work. For each light curve, this table shows the date of 
acquisition, the used instrument and filter, the exposure time $T_{\rm exp}$, the number of data points, the baseline function selected for 
our global analysis (see Sec.~\ref{sec:analysis}), the error correction factor $CF$ used in our global analysis, and the nature
of the eclipse (transit or occultation) For the baseline function, $p(\epsilon^N)$ denotes, respectively, a $N$-order polynomial function of
 time ($\epsilon=t$), the logarithm of time ($\epsilon=l$), $x$ and $y$ positions ($\epsilon=xy$), full-width at half-maximum FWHM ($\epsilon=fwhm$)
 and  FWHM in the $x$-direction ($\epsilon=fwhm_x$); $o$ denotes an offset at the time of a meridian flip of TRAPPIST (see \citealt{Gillon:2012fj}); BM 
 denotes the use of the BLISS-mapping technique \citep{Stevenson:2012vn} to improve the modeling of the position effects (see \citealt{Gillon:2014ix}).  }\label{tab:phot}
\end{center}
\end{table*}

\begin{figure*}  
\begin{center}  
	\begin{subfigure}[b]{0.45\textwidth}
		\includegraphics[width=\textwidth]{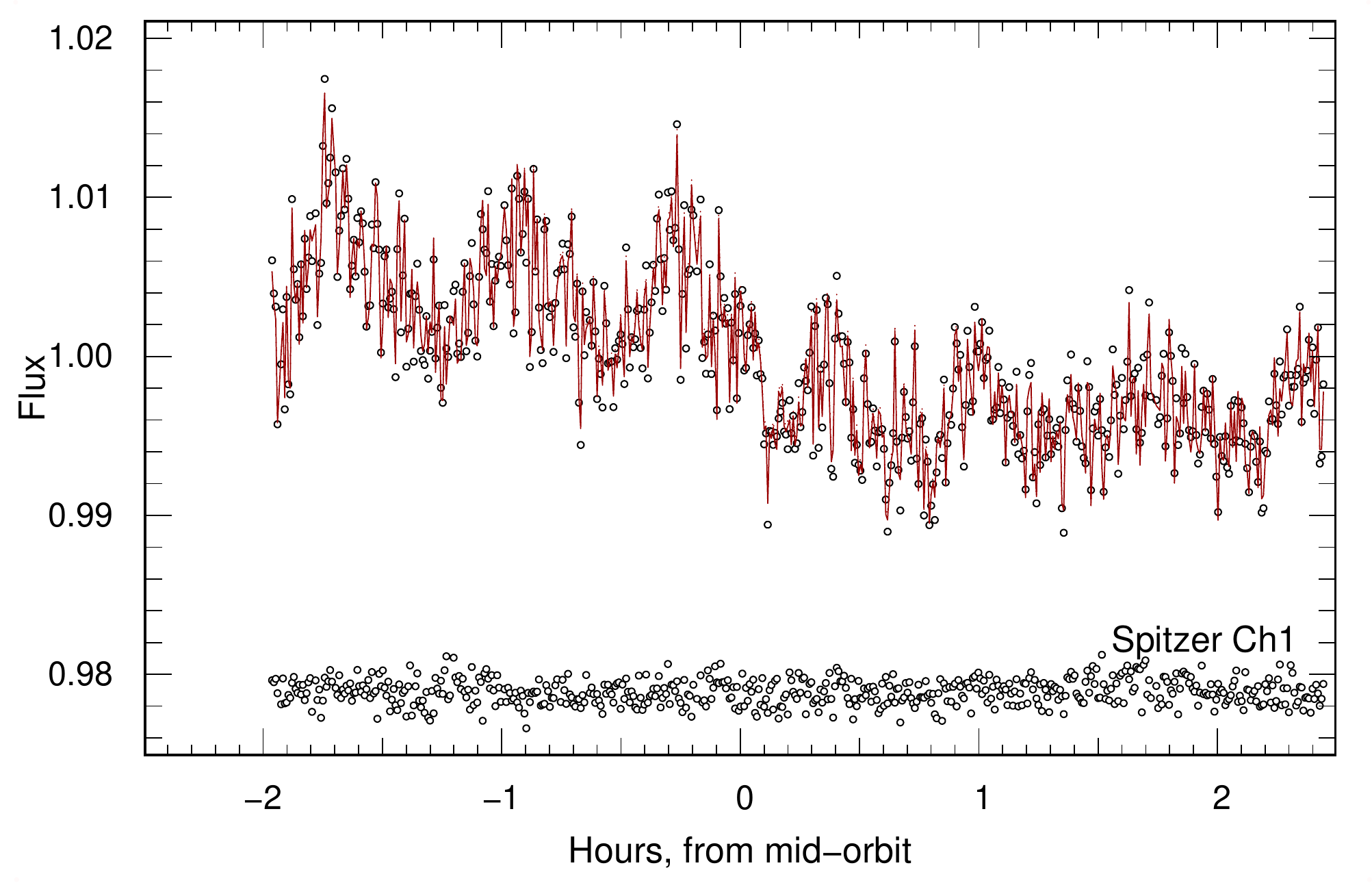}  
		\label{}  
	\end{subfigure}
	\begin{subfigure}[b]{0.45\textwidth}
		\includegraphics[width=\textwidth]{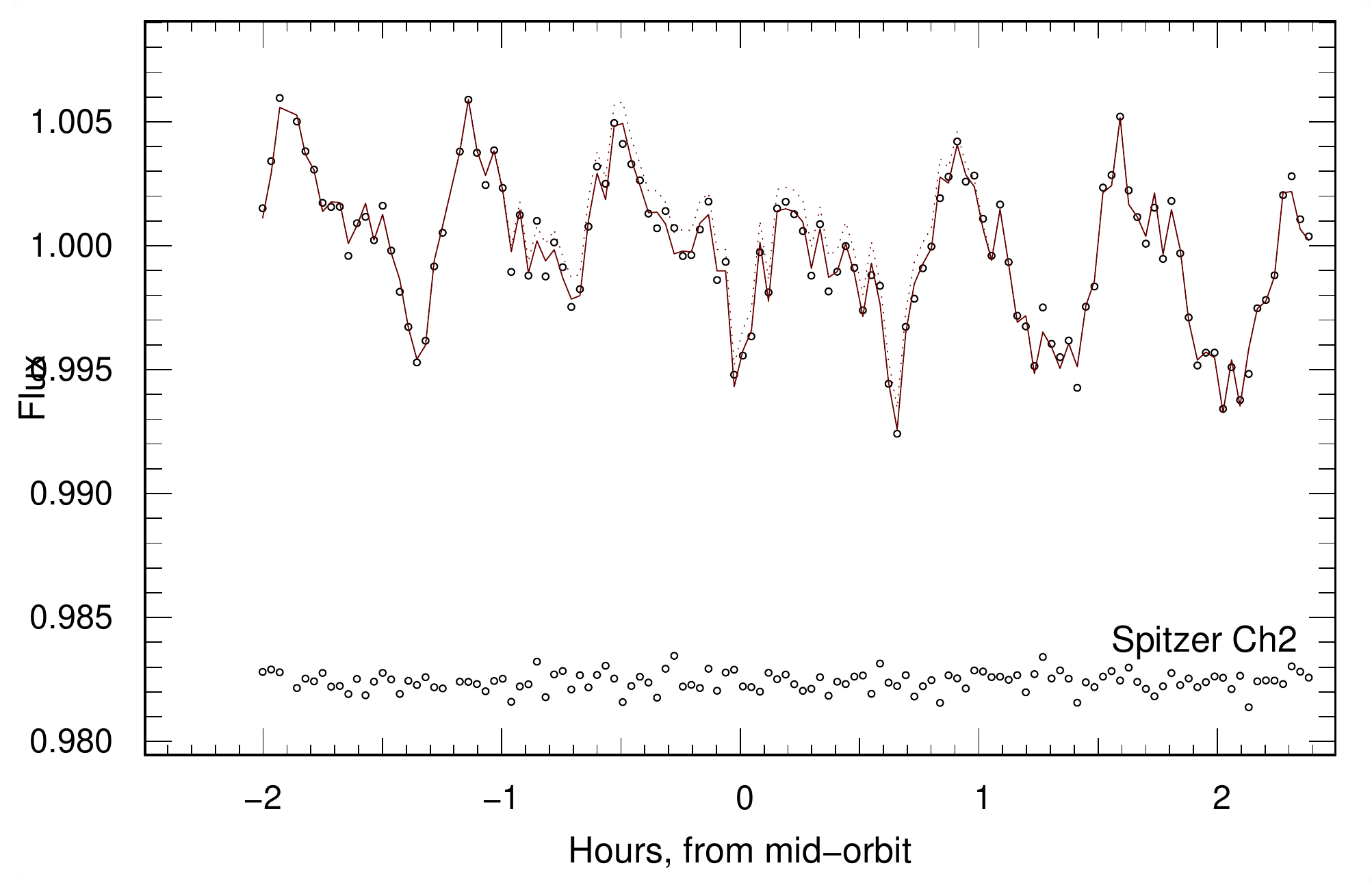}  
		\label{}  
	\end{subfigure}
	\begin{subfigure}[b]{0.45\textwidth}
		\includegraphics[width=\textwidth]{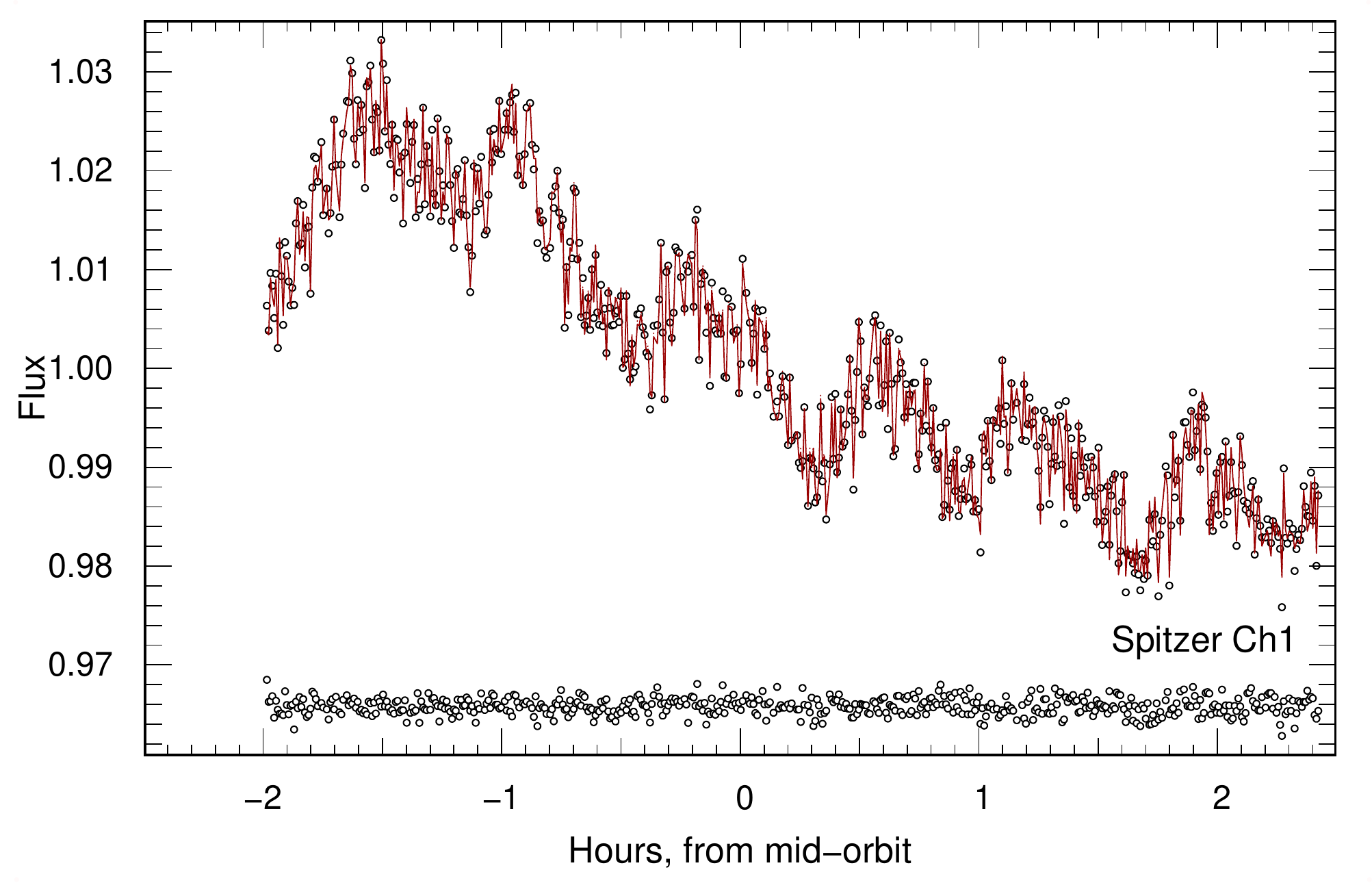}  
		\label{}  
	\end{subfigure}
	\begin{subfigure}[b]{0.45\textwidth}
		\includegraphics[width=\textwidth]{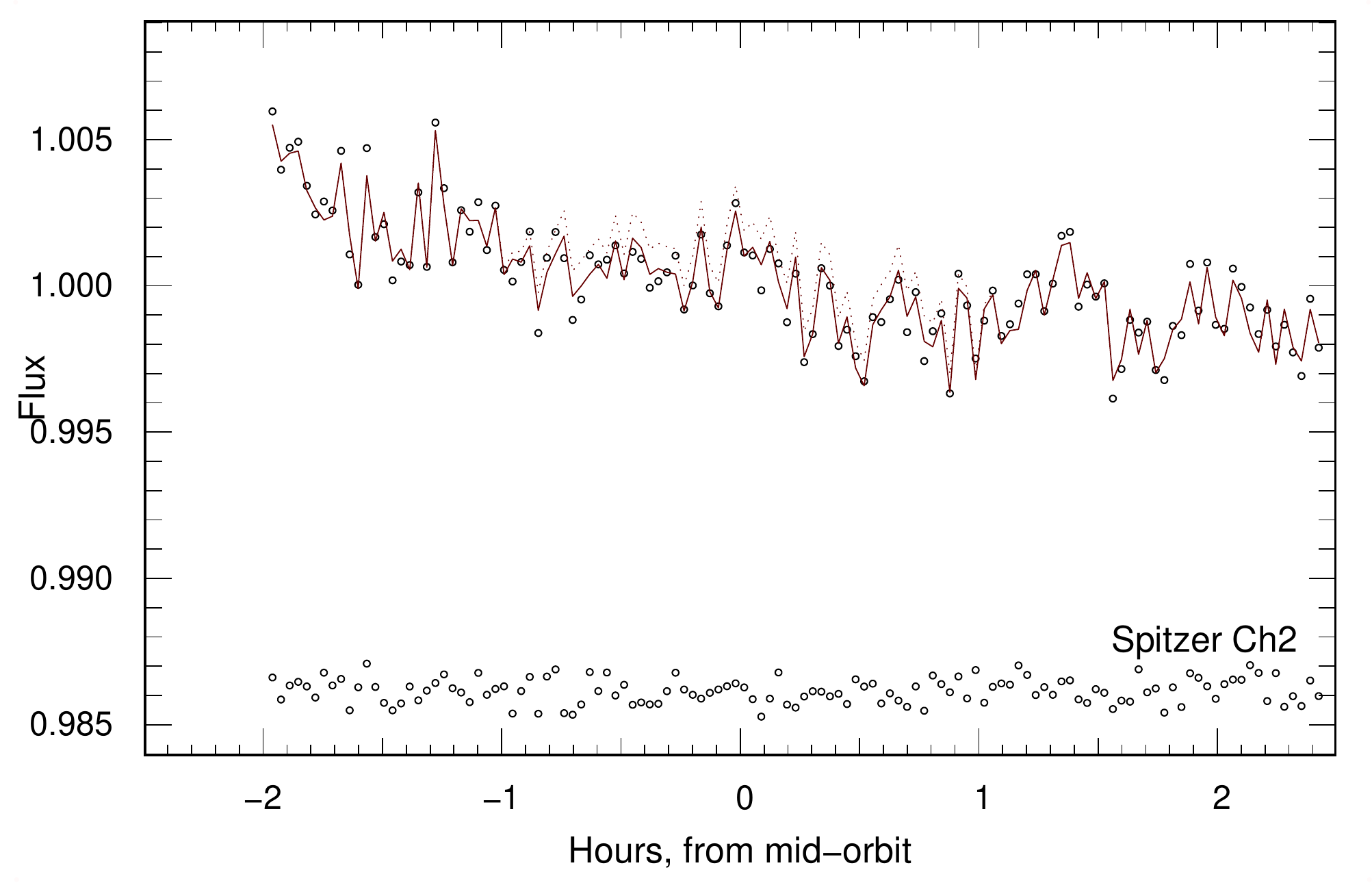}  
		\label{}  
	\end{subfigure}
	\caption{Photometry at occultation, as extracted from the {\it Warm Spitzer} frames, in both channels, with the initial ramp removed. The full model to the data is shown as a plain line; the residuals are displayed underneath. The corrections, including the variations in intra-pixel sensitivity, are isolated and drawn as a dotted line (barely noticeable here due to the weakness of the occupation signal compared to the corrections).
}\label{fig:raw_trans}  
\end{center}  
\end{figure*} 


\begin{figure*}  
\begin{center}  
	\begin{subfigure}[b]{0.45\textwidth}
		\includegraphics[width=\textwidth]{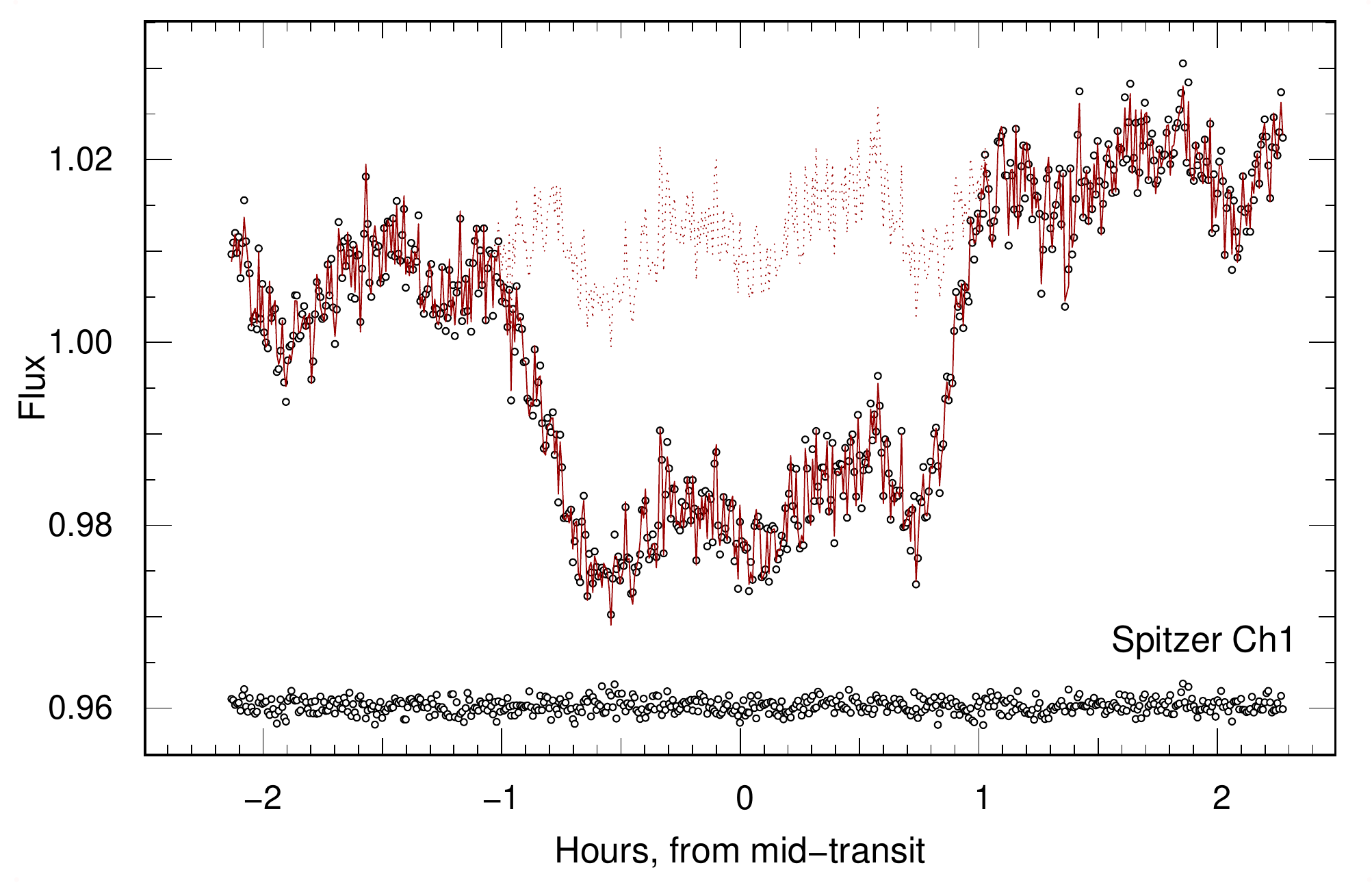}  
		\label{}  
	\end{subfigure}
	\begin{subfigure}[b]{0.45\textwidth}
		\includegraphics[width=\textwidth]{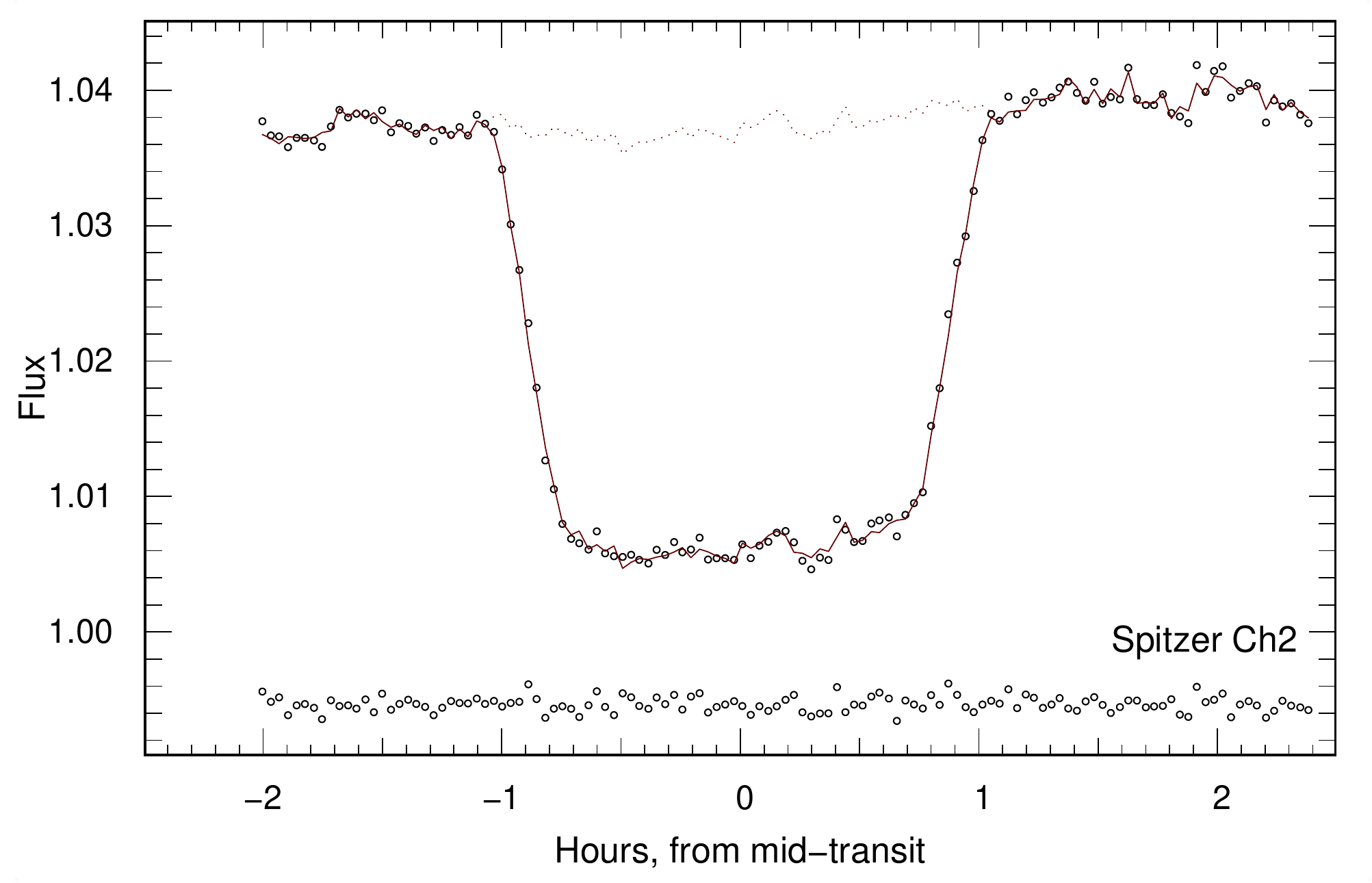}  
		\label{}  
	\end{subfigure}
	\caption{Photometry at transit, as extracted from the {\it Warm Spitzer} frames, in both channels, with the initial ramp removed. The full model to the data is shown as a plain line; the residuals are displayed underneath. The corrections, including the variations in intra-pixel sensitivity, are isolated and drawn as a dotted line.
}\label{fig:raw_occ}  
\end{center}  
\end{figure*} 

\begin{figure*}  
\begin{center}  
	\begin{subfigure}[b]{0.33\textwidth}
		\includegraphics[width=\textwidth]{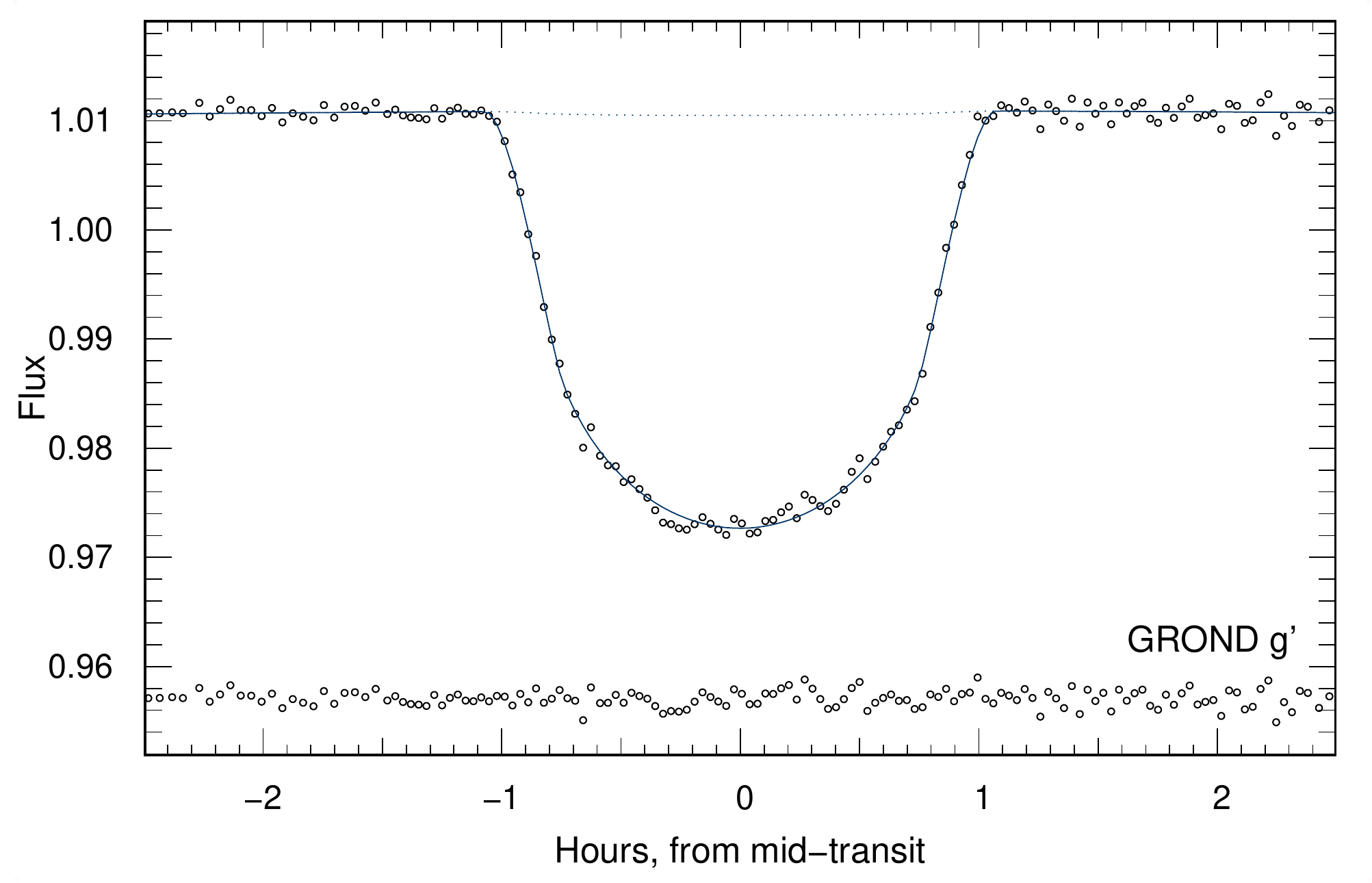}  
		\label{}  
	\end{subfigure}
	\begin{subfigure}[b]{0.33\textwidth}
		\includegraphics[width=\textwidth]{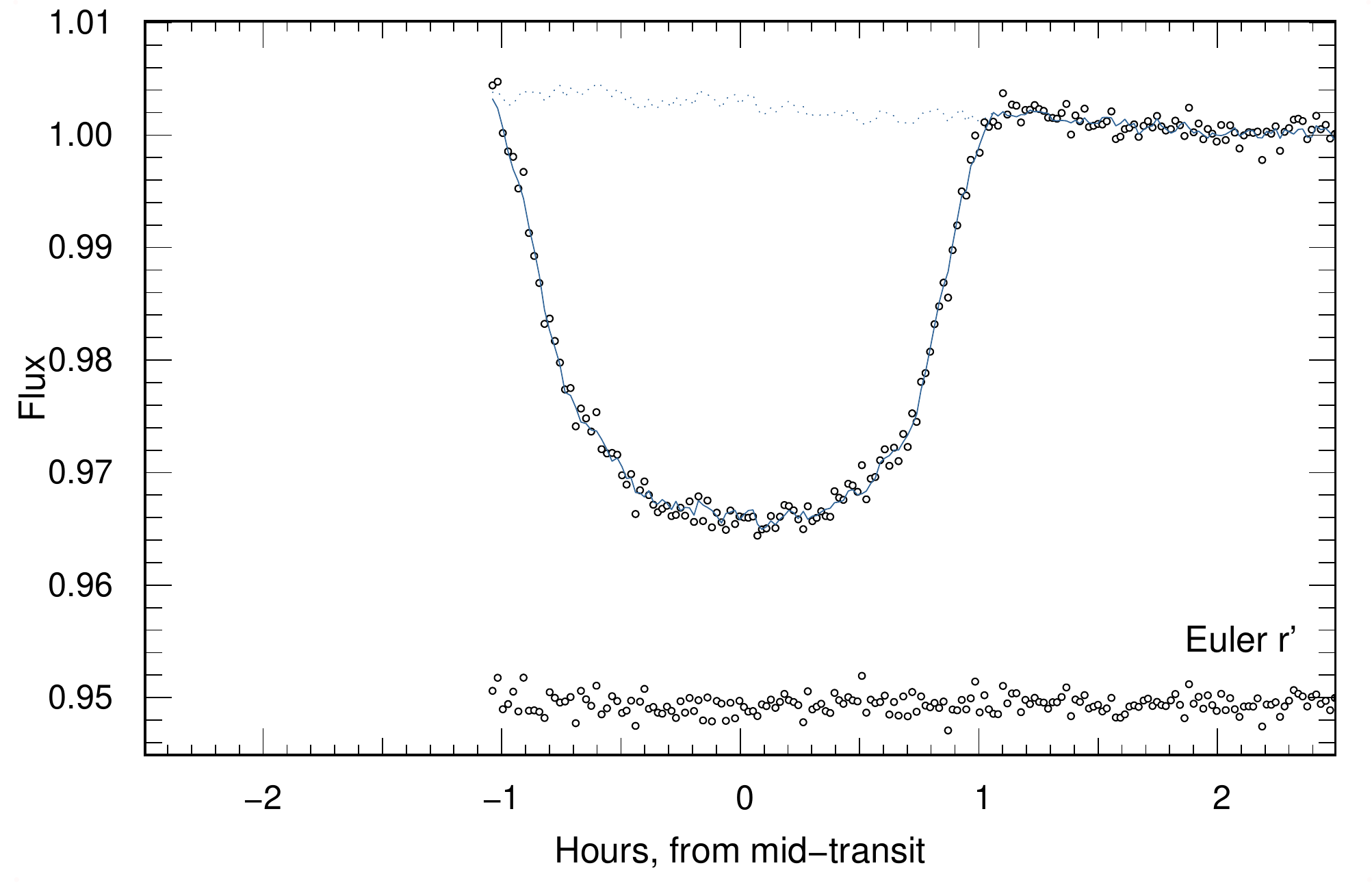}  
		\label{}  
	\end{subfigure}
	\begin{subfigure}[b]{0.33\textwidth}
		\includegraphics[width=\textwidth]{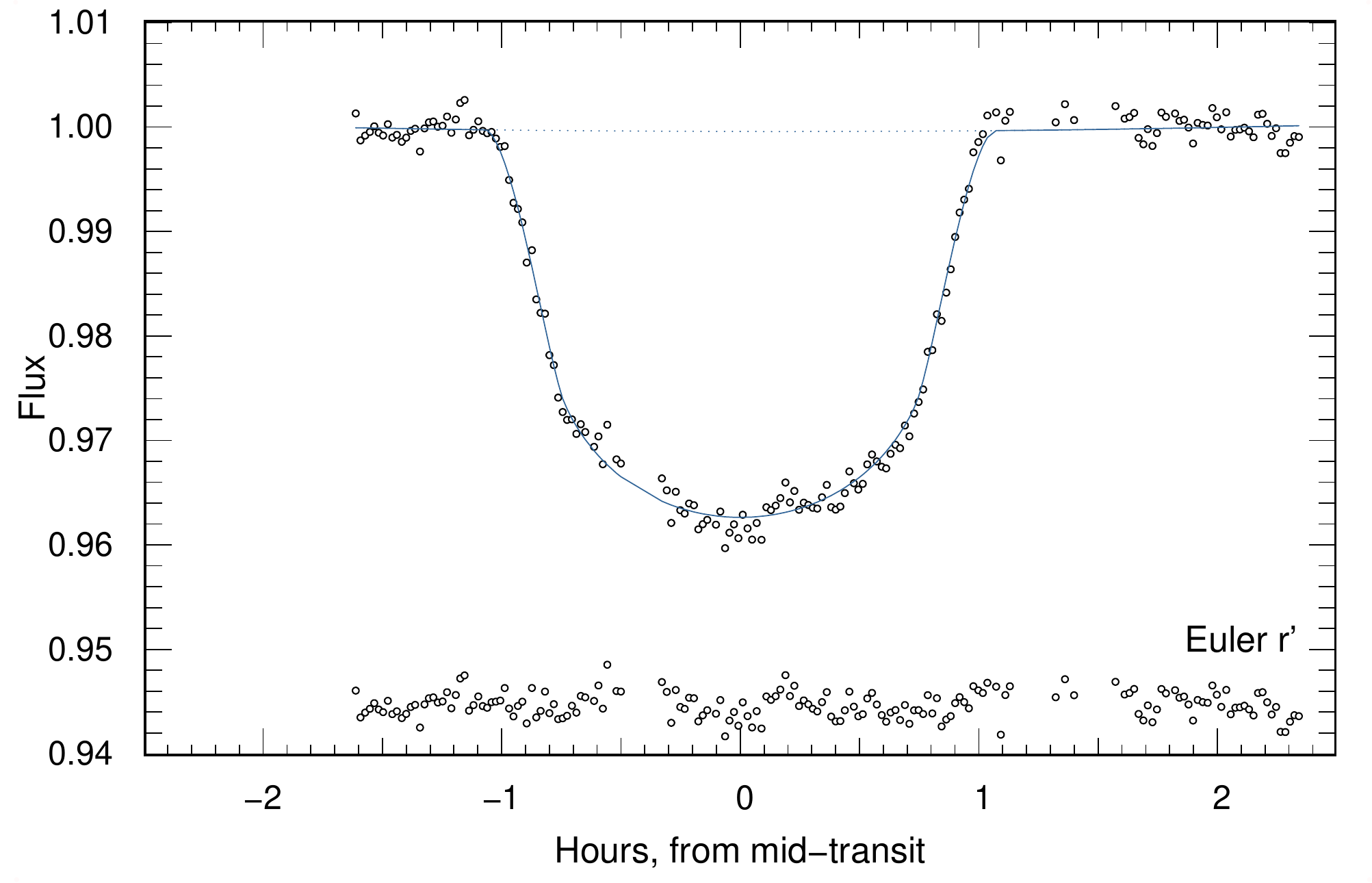}  
		\label{}  
	\end{subfigure}
	
	\begin{subfigure}[b]{0.33\textwidth}
		\includegraphics[width=\textwidth]{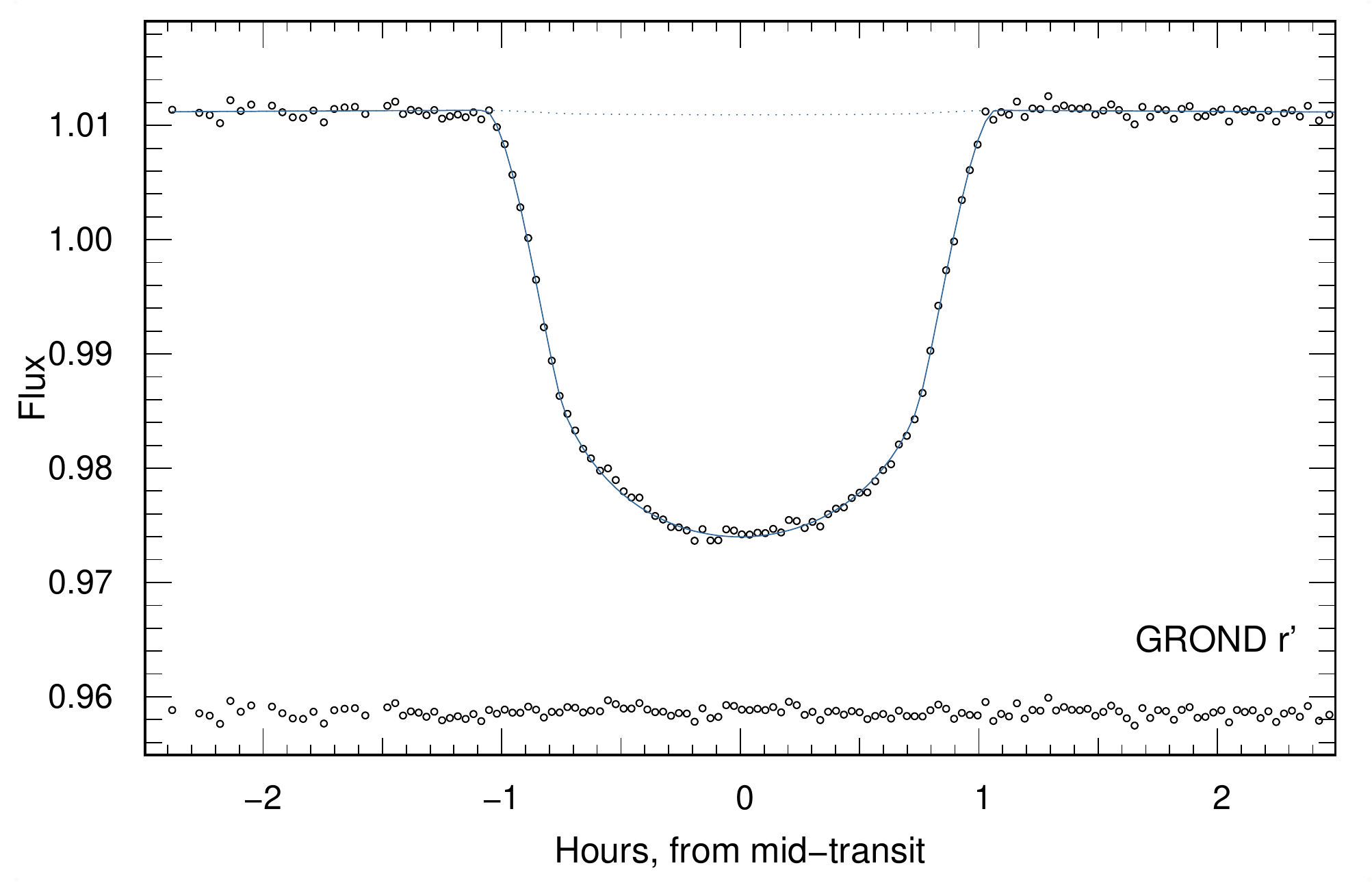}  
		\label{}  
	\end{subfigure}
	\begin{subfigure}[b]{0.33\textwidth}
		\includegraphics[width=\textwidth]{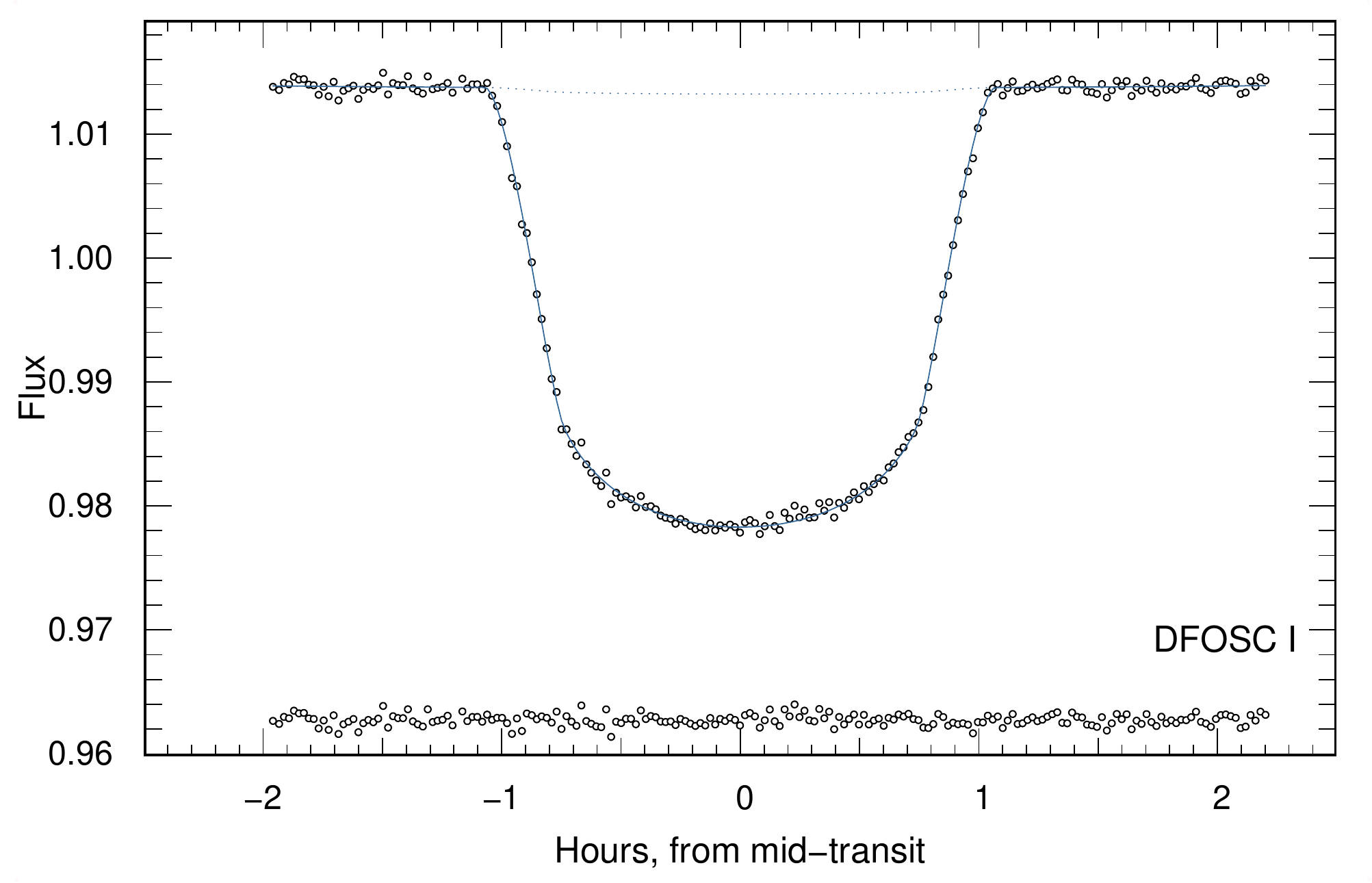}  
		\label{}  
	\end{subfigure}
	\begin{subfigure}[b]{0.33\textwidth}
		\includegraphics[width=\textwidth]{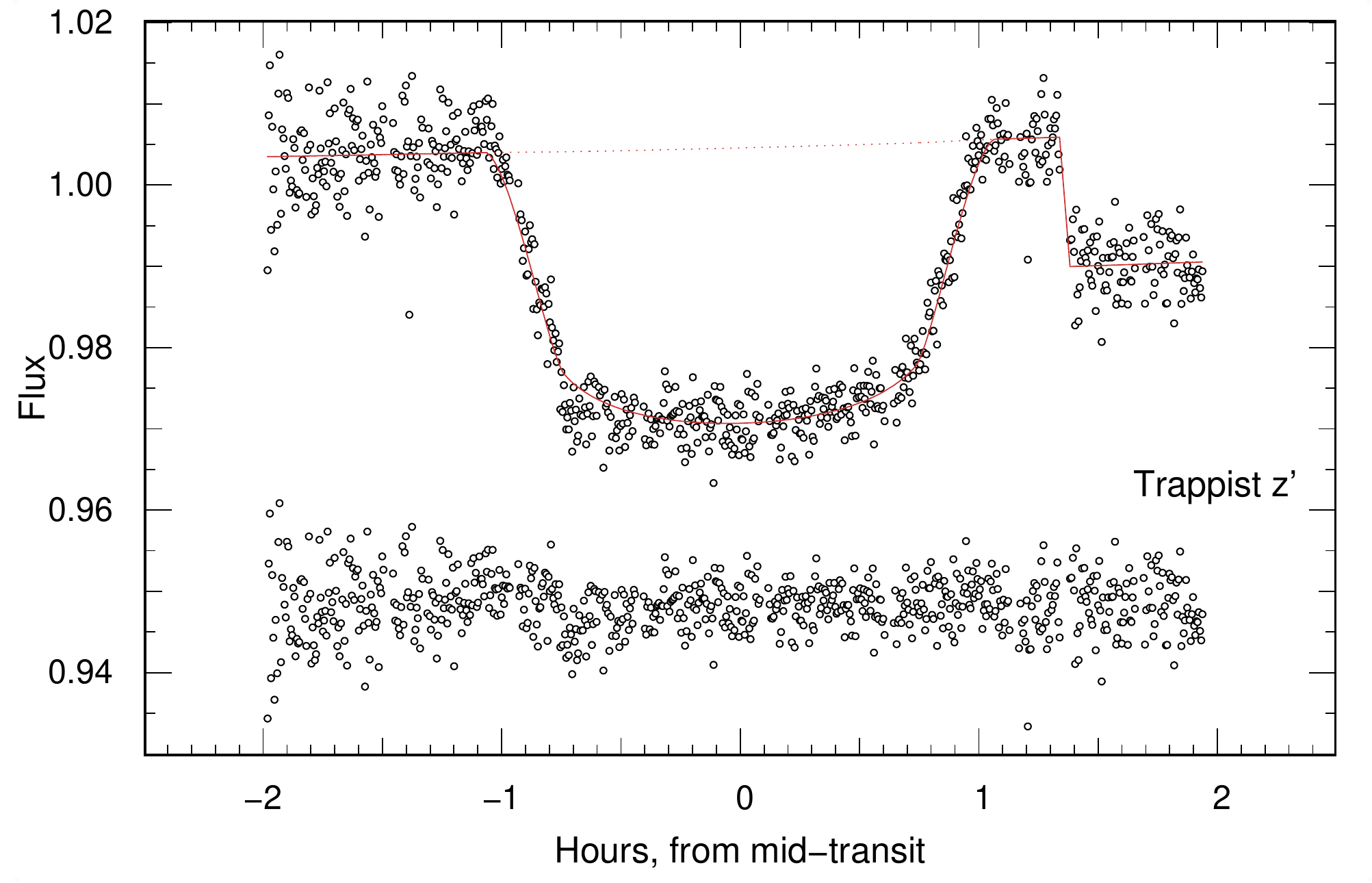}  
		\label{}  
	\end{subfigure}

	\begin{subfigure}[b]{0.33\textwidth}
		\includegraphics[width=\textwidth]{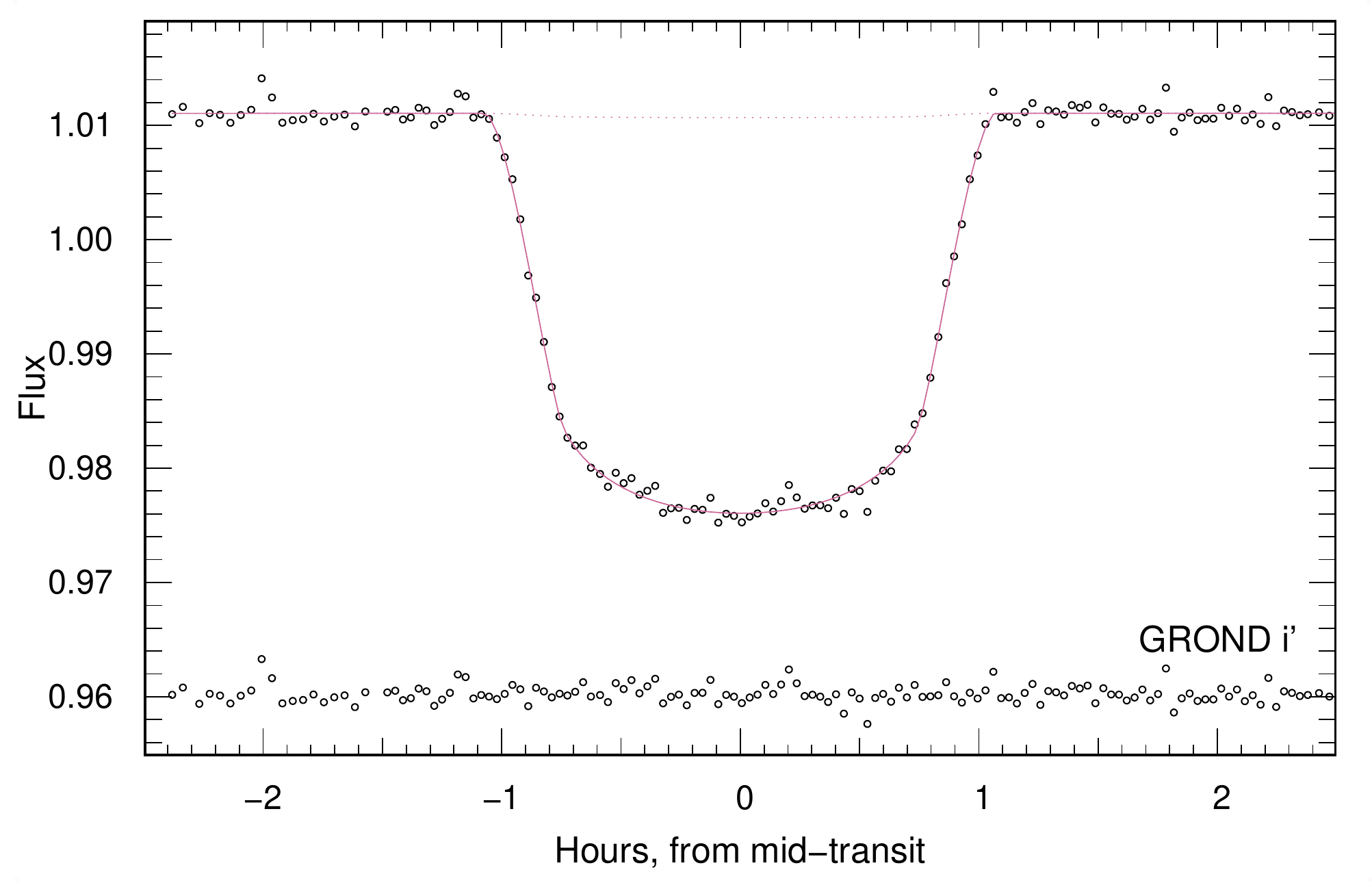}  
		\label{}  
	\end{subfigure}
	\begin{subfigure}[b]{0.33\textwidth}
		\includegraphics[width=\textwidth]{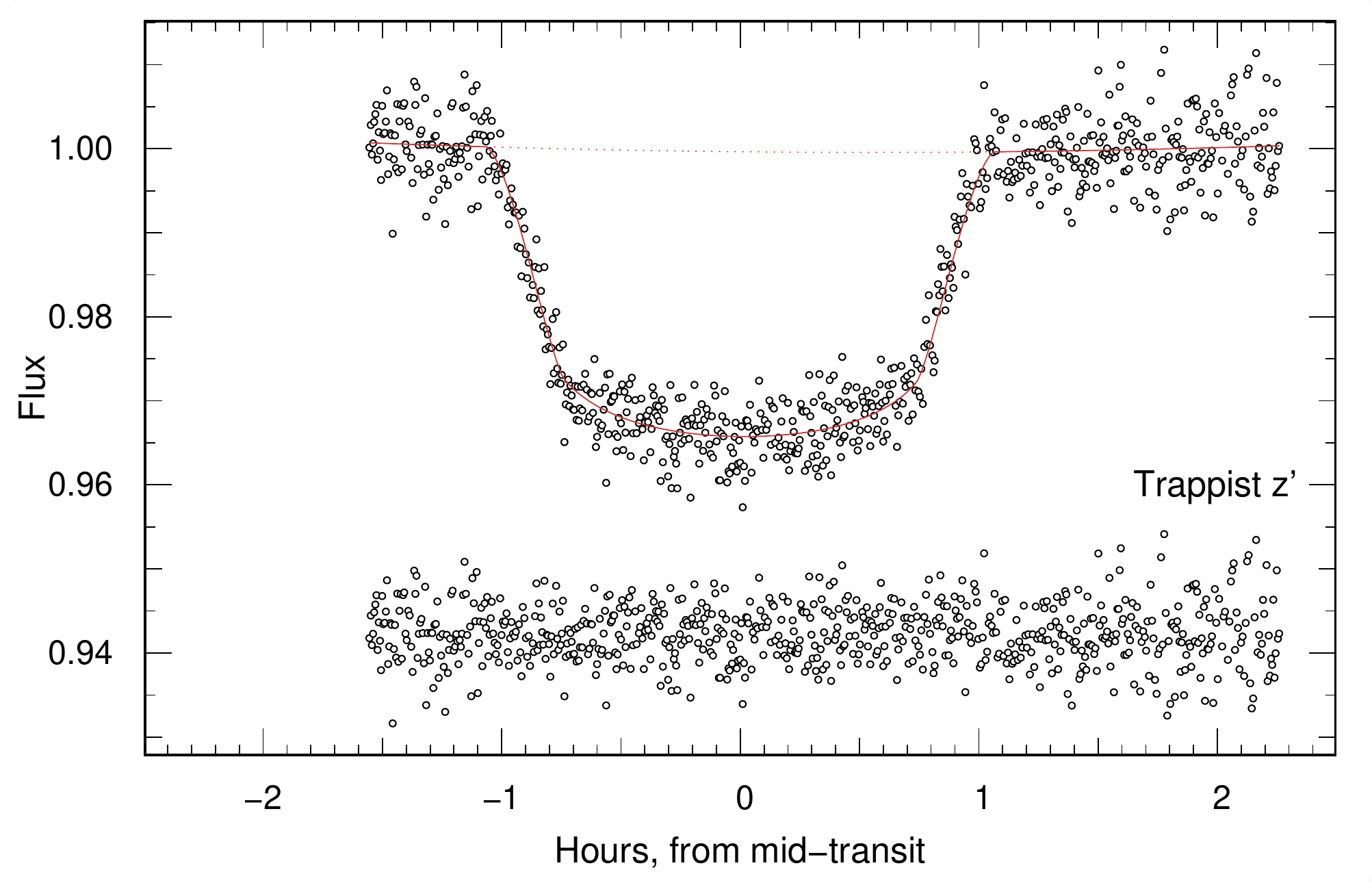}  
		\label{}  
	\end{subfigure}
	\begin{subfigure}[b]{0.33\textwidth}
		\includegraphics[width=\textwidth]{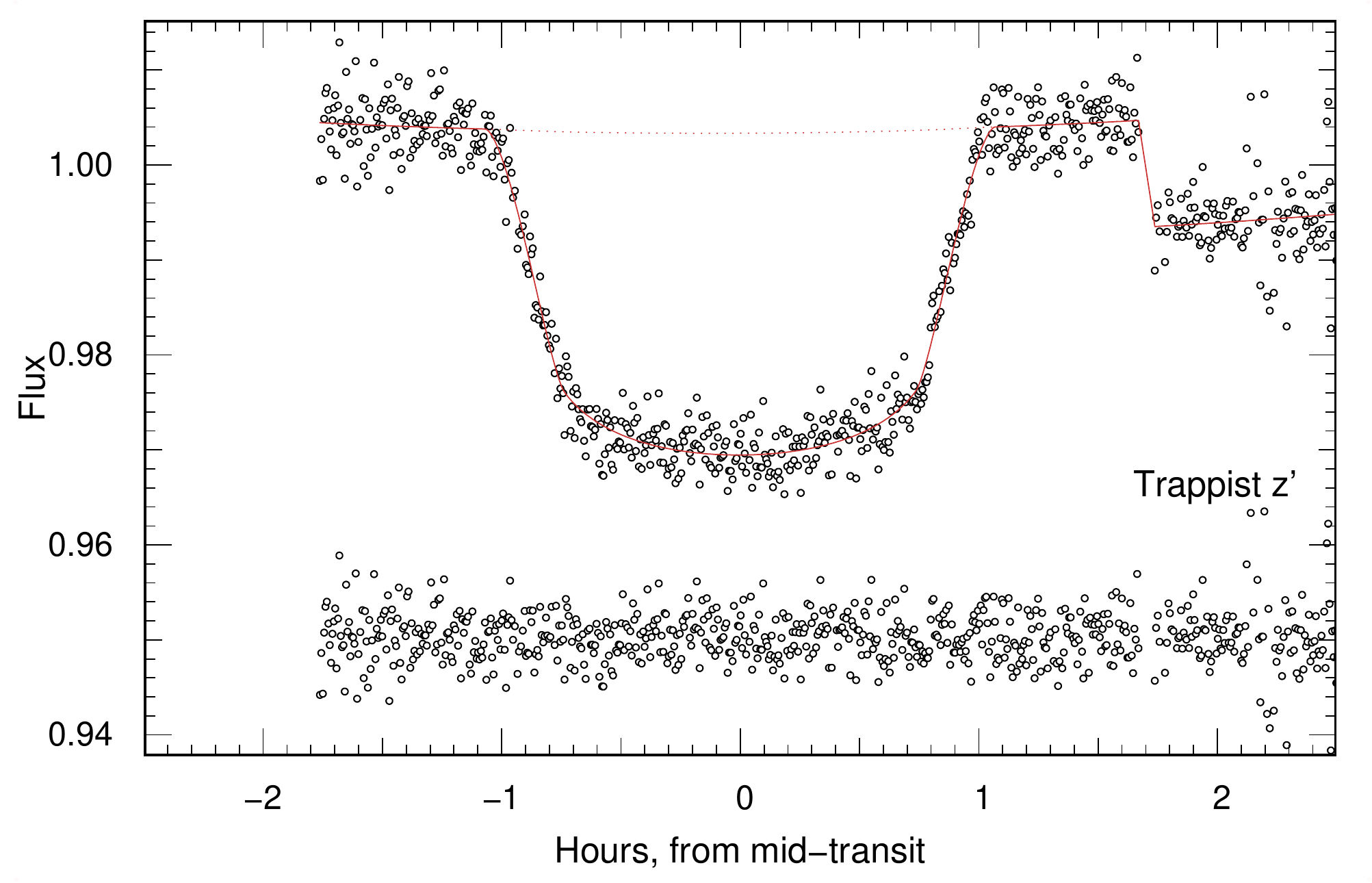}  
		\label{}  
	\end{subfigure}

	\begin{subfigure}[b]{0.33\textwidth}
		\includegraphics[width=\textwidth]{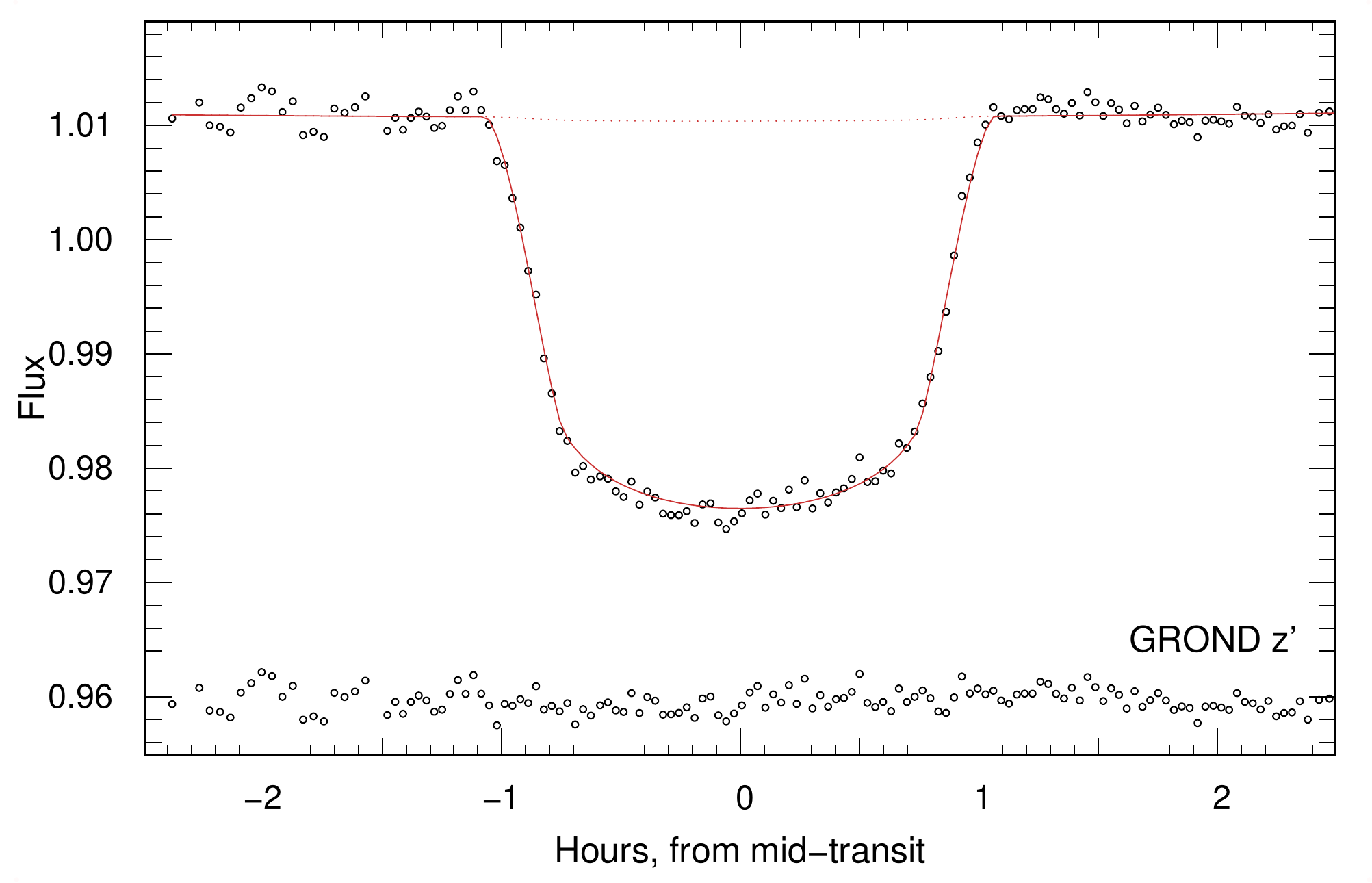}  
		\label{}  
	\end{subfigure}
	\begin{subfigure}[b]{0.33\textwidth}
		\includegraphics[width=\textwidth]{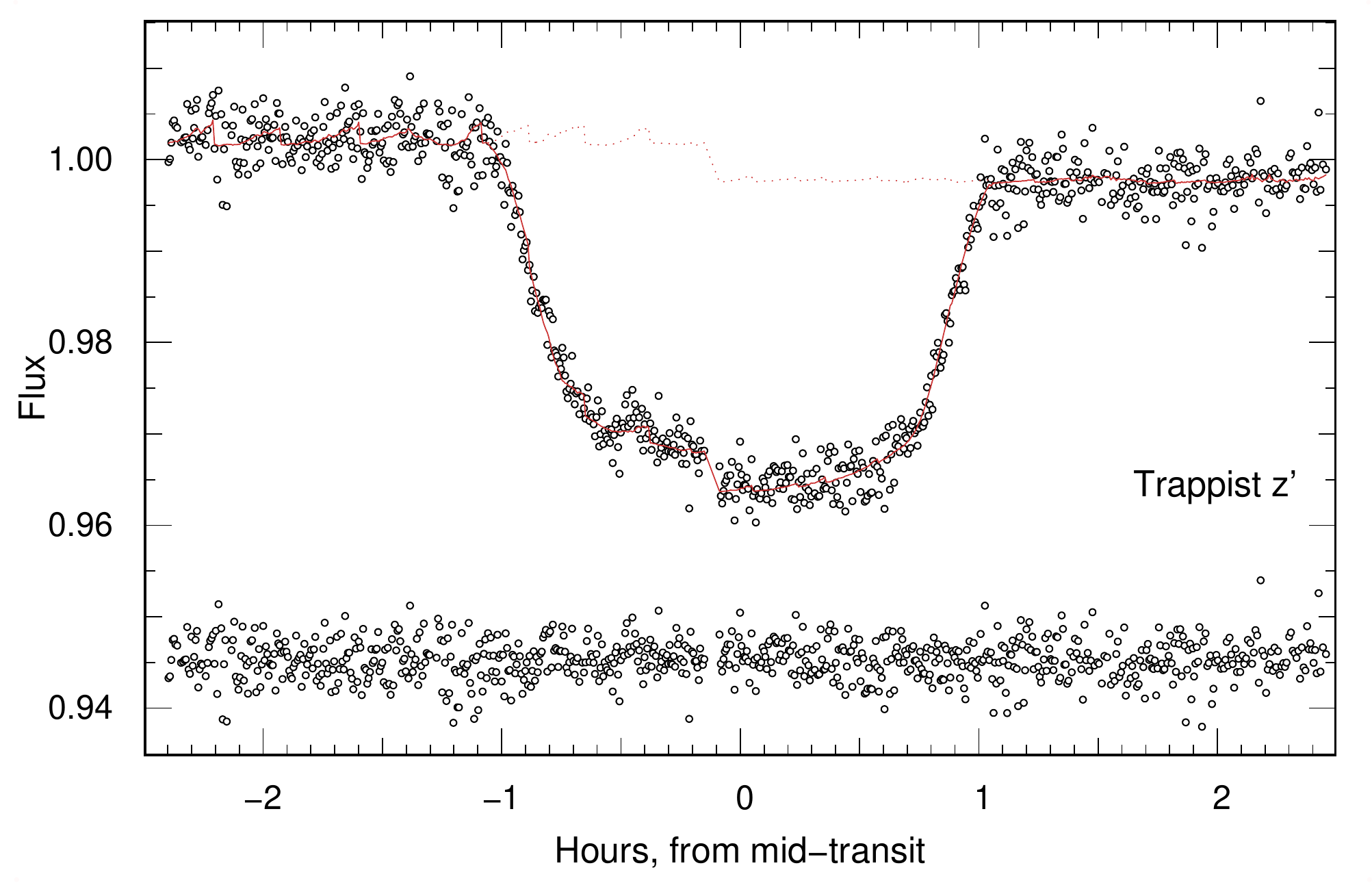}  
		\label{}  
	\end{subfigure}
	\begin{subfigure}[b]{0.33\textwidth}
		\includegraphics[width=\textwidth]{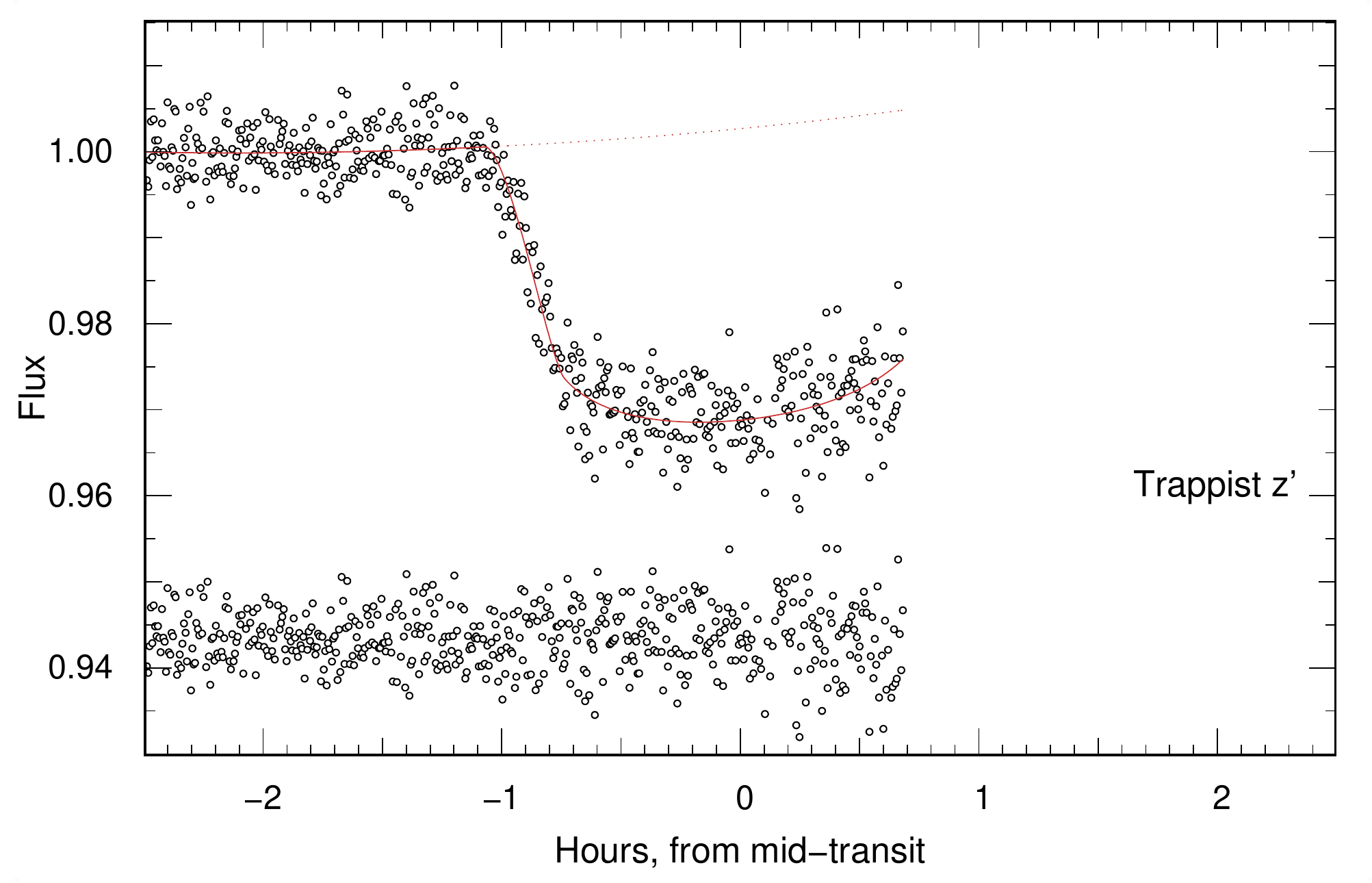}  
		\label{}  
	\end{subfigure}

	\caption{Photometry at transit, in the visible wavelengths. The full model to the data is shown as a plain line; the residuals are displayed underneath. The corrections, for instance airmass, or changes in seeing or pointing, are isolated and drawn as a dotted line.
}\label{fig:ground}  
\end{center}  
\end{figure*} 

\newpage

\bsp

\label{lastpage}

\end{document}